\documentclass[aps,prd,twocolumn,letterpaper,superscriptaddress,floatfix,nofootinbib]{revtex4-1}
\usepackage{amsmath,amssymb}

\usepackage{tikz}
\usepackage{tikz-feynman}
\usepackage[utf8]{inputenc}
\usepackage[english]{babel}
\usepackage{amsmath,amssymb,amsbsy,amstext,amsthm,booktabs,simplewick,exscale,relsize,slashed,graphicx,amsfonts,upgreek, xcolor, colortbl}
\usepackage{multirow,color,dcolumn,bm,enumerate}
\usepackage{hyperref}
\usepackage{feynmp-auto,axodraw2,tikz}
\usepackage{ulem}
\usepackage{comment}
\usepackage{amsmath}
\usepackage{ytableau}
\usepackage{braket}
\usepackage{booktabs}
\usepackage{graphicx}
\usepackage{dcolumn}
\usepackage{bm}
\usepackage{hyperref}
\usepackage{xcolor}
\usepackage{amsmath,amsfonts,amssymb,amsthm}
\DeclareUnicodeCharacter{2212}{-}

\usepackage{hyperref}
\usepackage[capitalise,noabbrev,nameinlink]{cleveref}

\usepackage{mathtools}

\newcommand{\looponly}[1]{\boldsymbol{#1}}              
\newcommand{\mixedchirality}[1]{\underline{\vphantom{y}#1}}
\newcommand{\bothfeatures}[1]{\underline{\vphantom{y}\boldsymbol{#1}}}

\begin{document}

\allowdisplaybreaks[1]
\title{Energy-Enhanced Expansion of the Standard Model Effective Field Theory}
\author{Benoît Assi}
\affiliation{Department of Physics, University of Cincinnati, Cincinnati, OH, 45221, USA}
\affiliation{Particle Theory Department, Fermilab, Batavia, IL, 60510,  USA}
\email{bassi@fnal.gov}
\author{Adam Martin}
\affiliation{Department of Physics, University of Notre Dame, South Bend, IN, 46556 USA}
\email{amarti41@nd.edu}
\date{\today}

\begin{abstract}
We formalize energy-scaling arguments in the Standard Model Effective Field Theory (SMEFT) to estimate effects of operators up to dimension ten. Our approach relies on weakly-coupled UV completions with no presumed large hierarchies between Wilson coefficients. We introduce a classification based on the number of external legs and an energy-counting parameter, we establish a dual expansion in \(v/\Lambda\) and \(E/\Lambda\). Extending to four-, five-, and six-particle vertices, our framework highlights energy-enhanced operators that dominate high-energy processes at the HL-LHC. This organization streamlines experimental analyses to only include operators with energetic impact in their analyses and enhances the discoverability of new physics within the SMEFT framework.
\end{abstract}

\maketitle

\section{Introduction}
\label{sec:intro}

The Standard Model Effective Field Theory (SMEFT) provides a robust framework for systematically extending the Standard Model (SM) to include higher-dimensional operators~\cite{Buchmuller:1985jz, Grzadkowski:2010es, Brivio:2017vri}. These operators encode the effects of heavy physics, offering predictions beyond the SM within a controlled expansion in powers of $\frac{1}{\Lambda}$, where $\Lambda$ is the scale of new physics. 
To effectively assess the impact of these operators at high energies, we introduce an energy-based power counting scheme and classification to quantify the degree of enhancement one can expect from SMEFT operators up to dimension ten.

A wide range of processes—including precision electroweak tests, Higgs phenomenology, and top-quark dynamics—has been extensively studied in SMEFT~\cite{Araz:2020zyh,Contino:2013kra,Falkowski:2014tna,Englert:2014cva, Gupta:2014rxa, Amar:2014fpa, Buschmann:2014sia, Craig:2014una, Ellis:2014dva, Ellis:2014jta, Banerjee:2015bla, Englert:2015hrx, Contino:2016jqw, Biekotter:2016ecg, Barklow:2017suo, Barklow:2017awn, Englert:2017aqb, banerjee1, Grojean:2018dqj,Biekotter:2018rhp, Goncalves:2018ptp,Gomez-Ambrosio:2018pnl, Freitas:2019hbk,Banerjee:2019pks, Banerjee:2019twi, Biekotter:2020flu}. Within the framework of the SMEFT, the squared amplitude for a generic process can be analytically described by the following expansion:
\begin{align}
|\mathcal{A}|^2 = |A_{\text{SM}}|^2 &\Bigg\{ 1 + \frac{2\, \text{Re}(A_{\text{SM}}^* A_6)}{\Lambda^2 |A_{\text{SM}}|^2}
\nonumber \\& 
+ \frac{1}{\Lambda^4} \left( \frac{|A_6|^2}{|A_{\text{SM}}|^2} + \frac{2\, \text{Re}(A_{\text{SM}}^* A_8)}{|A_{\text{SM}}|^2} \right) + \cdots \Bigg\},
\end{align}
where \( A_6 \) and \( A_8 \) represent contributions from dimension-six and dimension-eight operators, respectively, and \( \Lambda \) denotes the scale of new physics. By factoring out the SM amplitude squared, all terms within the braces must be dimensionless. The scale \( \Lambda \) organizes the SMEFT expansion, necessitating compensation with powers of relevant dimensionful scales in the problem, such as the Higgs vacuum expectation value \( v \) or the characteristic energy scale \( E \) of the process.

Current global bounds on SMEFT operators are predominantly constrained by precision measurements of well-studied, resonance-dominated processes such as \( Z \to \ell\ell \) and \( gg \to h \to \gamma\gamma \)~\cite{Ellis:2018gqa,Ellis:2020unq,Bartocci:2023nvp,Giani:2023gfq}. These processes primarily probe the \( v/\Lambda \) expansion due to their resonant nature. As the High Luminosity Large Hadron Collider (HL-LHC) progresses and collects more data, the focus will shift towards non-resonant processes, including multi-boson production (e.g., \( pp \to WWW \), \( pp \to WZ \)) and high-energy scattering events (such as \( pp \to t\bar{t}h \) and \( pp \to ZZ \)), which fall into the category of precision measurements sensitive to the \( E/\Lambda \) expansion.  In particular, although events in the kinematic tails—characterized by high transverse momentum or large invariant masses—are statistically suppressed, organizing operators by their energy scaling isolates the subset whose contributions grow with energy, thereby sharpening the sensitivity to BSM effects and enabling more stringent tests of the SMEFT framework. 

While the increased sensitivity to \( E/\Lambda \) opens new avenues for the discovery of physics beyond the SM, it also necessitates careful consideration to ensure the validity of the effective field theory (EFT) approach. Specifically, the expansion in \( E/\Lambda \) must remain controlled, avoiding scenarios where higher-dimensional operators could invalidate the perturbative expansion or the EFT's applicability within the energy regime of interest. To accommodate the diverse kinematic configurations encountered in future SMEFT studies, it is advantageous to adopt an organization of SMEFT effects that systematically accounts for both \( v/\Lambda \) and \( E/\Lambda \) power counting. In this work, we propose to classify SMEFT operators based on the number of external legs in each vertex and introduce an additional parameter \( \lambda \) in the Lagrangian to count powers of energy systematically. This vertex-based classification, as opposed to the traditional field-content-based approach, allows for a clearer separation of contributions according to their energy scaling, thereby streamlining both theoretical analyses and experimental searches.

From an experimental standpoint, this organization significantly reduces the number of operators that need to be considered, as operators exhibiting energy enhancement will dominate the kinematic tails of relevant distributions. Furthermore, additional assumptions, such as specific flavor structures or the origin of operator matching (tree-level versus loop-level), can further constrain the operator list. The approach we advocate for here not only minimizes theoretical clutter but also ensures that experimental analyses remain sensitive to meaningful deviations without being diluted by operators that do not contribute significantly to the processes under study.

From the perspective of model building and theoretical investigations, our classification highlights the characteristics of new physics models that render them more accessible to discovery at the HL-LHC. Specifically, models that generate energy-enhanced operators, particularly those arising from tree-level matching conditions, tend to be more discoverable within this framework. One can then combine our operator counting with explicit vertex-generation automated tools such as Refs.~\cite{Dedes:2017zog,Dedes:2023zws}, to immediately translate into full amplitudes ready for experimental fits.
Importantly, these criteria do not alter the SMEFT in any way but rather provide a systematic organizational framework. 

Moreover, we offer a systematic way to manage the rapid proliferation of operators that arise at higher mass dimensions~\cite{Henning:2015alf}. This focused approach is especially beneficial in precision phenomenology, where combining data from different energy scales necessitates careful renormalization group (RG) evolution~\cite{Assi:2023zid,Chala:2021pll, DasBakshi:2022mwk, Bakshi:2024wzz, Boughezal:2024zqa,Helset:2022pde}. In effect, by concentrating on the most promising operators, the overall theoretical framework becomes more tractable, allowing for a clearer connection between experimental observations and the underlying EFT. Indeed, our vector‑boson fusion (VBF) analysis at the HL‑LHC~\cite{Assi:2024zap} is a prime example of how this discoverability‑focused strategy can enhance sensitivity and provide a clear link between experimental observables and the underlying EFT contributions. While Ref.~\cite{Assi:2024zap} focused on VBF Higgs production, other studies have successfully applied similar ideas to different processes~\cite{Kim:2022amu,Martin:2023tvi,Corbett:2023yhk,Degrande:2023iob}.

Building on the results of Ref.~\cite{Helset:2020yio}, which organized operators by their \(v/\Lambda\) scaling for 2‑ and 3‑point interactions, we now extend the classification by introducing an explicit \(E/\Lambda\) power counting applicable to vertices with any number of legs. In particular, we systematically investigate vertices with up to six external legs. This work is organized as follows: In Section~\ref{sec:energy_expansion} we introduce the energy–enhanced expansion of the SMEFT and establish the associated power counting. In Section~\ref{sec:further_enhancements} we extend the energy classification to incorporate additional enhancements employed in our examples. In Section~\ref{sec:energy_scaling_tables} we provide energy scaling tables for operator classes up to dimension ten. Section~\ref{sec:examples} applies our framework to several representative 4-, 5-, and 6-point processes, and in Section~\ref{sec:discussion} we discuss the implications of our results and conclude.

\section{Energy Expansion}
\label{sec:energy_expansion}

We begin with the SMEFT Lagrangian:
\begin{equation}
    \mathcal{L}_\text{SMEFT} = \mathcal{L}_\text{SM} + \sum_{i, j} \frac{c_j^{(i)}}{\Lambda^{i}} \mathcal{O}_j^{(4+i)},
    \label{eq:SMEFT_Lagrangian}
\end{equation}
where $i$ indexes the mass dimension above four, and $c_j^{(i)}$ and $\mathcal{O}_j^{(4+i)}$ represent the Wilson coefficients and corresponding operators of dimension $4+i$, respectively.  In the following we assume that \(c_j^{(i)}/\Lambda^{i}\) are all of comparable magnitude—an assumption that reflects a weakly coupled new physics scenario. In the absence of a specific UV model that enforces a strong hierarchy (for example, via a symmetry or dynamical mechanism), there is no a priori reason to expect that these coefficients should differ substantially. This is reasonable to assume since generically, matching calculations in perturbative UV completions tend to produce contributions of similar size across different operators \cite{deBlas:2014mba,deBlas:2017xtg,Carmona:2021xtq,Fuentes-Martin:2022jrf, Li:2023cwy}. If a specific UV scenario generates large hierarchies—e.g.\ strong
couplings, loop suppressions
$1/16\pi^{2}$, or symmetry-induced zeros—one simply multiplies the
$\lambda$ weight of each affected vertex by the corresponding
hierarchy factor.  
In that way our counting provide a baseline: strong or loop-suppressed coefficients do not invalidate the energy counting, rather they reshuffle which operator dominates at a given order.

The Lagrangian in Eq.~\eqref{eq:SMEFT_Lagrangian} post Electroweak Symmetry Breaking (EWSB)  can be ordered by the number of particles in a tree-level vertex. This is valid since not all operators will contribute to a vertex with a given number of legs; operators containing many fields will not contribute at tree level to vertices with few legs, while operators with only a few fields cannot  contribute to vertices with many legs.
Thus, for operators that can contribute to a given vertex leg multiplicity, we can derive how each scales with $v$ and $E$, the two scales in the problem.

To systematically organize the SMEFT Lagrangian based on the number of legs in each vertex, we propose the following decomposition:
\begin{equation}
    \mathcal{L}_\text{SMEFT}^{(n)} = \sum_j g_\text{SM}^j \lambda_j^{(n)} \mathcal{O}_\text{SM}^j + \sum_{i,k} c_k^{(i)} \frac{\lambda_k^{(n)}}{\Lambda^{i}} \mathcal{O}_k^{(4+i)}.
    \label{eq:SMEFT_n_point}
\end{equation}
where each component $\mathcal{L}_\text{SMEFT}^{(n)}$ contains interactions involving $n$ external legs. Here, we introduce a new power-counting parameter $\lambda^{(n)}$, which depends on the number of legs $n$ in a vertex, alongside the cutoff scale $\Lambda$, which depends on the EFT regime. Specifically, we define the scaling of $E$, $v$ and $\Lambda$ as
\begin{equation}
    (\Lambda, E, v) \sim
    \begin{cases}   
        (\lambda^{-3}, \lambda^{-2}, \lambda^{-1}) & \text{if } \Lambda \gg E \gg v, \\
        (\lambda^{-3}, \lambda^{-1}, \lambda^{-1}) & \text{if } \Lambda \gg E \sim v,
    \end{cases}
    \label{eq:scaling_conditions}
\end{equation}
for a dimensionless power counting parameter $\lambda\ll 1$ that is set by the field and derivative content of $\mathcal{O}$, as will be discussed below. Given that the size of $\Lambda$ is accounted for in this scheme, one can take it to be of order one in~\eqref{eq:SMEFT_n_point}.
This organization exploits the fact that operators can affect vertices with several different leg multiplicities. The scaling at the different multiplicities do not have to be the same, so $\lambda^{(n)} \ne \lambda^{(n')}$ in general.  

Moreover, we note here that to establish a consistent counting algorithm it is crucial to work in a minimal Warsaw-like basis. Our counting scheme treats each derivative as potentially contributing an energy factor. Given this limitation, eliminating redundant derivatives (those removable by integration by parts or via equations of motion) is essential to prevent an artificial boost in the counted energy scaling.

For 2 and 3- particle vertices, the energy classification can be read off immediately thanks to the geometric SMEFT organization. Specifically, due to the special kinematics of 2 and 3 particle vertices, Ref.~\cite{Helset:2020yio,Hays:2020scx,Corbett:2021eux} showed that the $E$ scaling is fixed by the lowest-dimension operator that contains the vertex, with even higher-dimensional operators only contributing factors of $v$. This organization also reduces the number of minimal operators to consider per vertex, as illustrated in Table~\ref{fig:geo_org}. For example, $(H^\dag H) X_{\mu\nu}X^{\mu\nu}$ generates a $hVV$ which scales as $\lambda = vE/\Lambda^2$. At dimension eight and higher, adding derivatives always results in operators that reduce by equations of motion and integration by parts to other terms, so our only option is $(H^\dag H)^n X_{\mu\nu}X^{\mu\nu}$, which scales as the original dimension six scaling times $(v^2/\Lambda^2)^{n-1}$. This is particularly useful because ``universal" corrections related to SM parameter input involve a small number of operators, and the simplest building blocks, three-point vertices, also involve a limited set of operators. Consequently, the bulk of the operators are relegated to four-point and higher interactions, making them more process-specific. 

The organization in Eq.~\eqref{eq:SMEFT_n_point} extends this energy classification concept to higher multiplicity ($\ge 4$ particle) vertices. As the kinematics is no longer trivial (momentum products do not just reduce to combinations of masses),  the $E$ scaling is not set by the lowest dimension operator contributing to a given vertex. While this means more complicated scalings are possible, and the scaling can change as we increase the mass dimension, it is straightforward to determine the scaling, at as we will show for dimension six through to dimension ten operators. This suffices for the purposes of identifying which effects through $\mathcal O(1/\Lambda^6)$ are the most likely to impact tails of kinematic distributions -- assuming that all Wilson coefficients are the same order of magnitude.

To determine the energy power-counting for a vertex, we first note that an on-shell $n$-point vertex has mass dimension $d = 4 - n$. The dimension of a SMEFT vertex, $\mathcal{V}^{(n)}$, is related to $v$, $E$, and $\Lambda$ by
\begin{equation}
    \left[\frac{E^q v^p}{\Lambda^D}\right] = q + p - D = d,
    \label{eq:dimension_relation}
\end{equation}
where $q$ and $p$ are the powers we want to determine and $D$ is the mass dimension of the operator inserted, given by
\begin{equation}
    D = \frac{3}{2} N_f + N_H + 2 N_X + N_D - 4.
    \label{eq:operator_dimension}
\end{equation}
where:
- \( N_{f} \): Number of fermions
- \( N_H \): Number of Higgs bosons
- \( N_X \): Number of external bosons
- \( N_D \): Number of covariant derivatives.

The algorithm is then as follows, given that the operator is able to generate at tree level the $n$-point vertex, meaning $N_f+N_X\leq n\leq 2N_f+N_X+N_D+N_H$, the vertex with the smallest power of $v$, denoted as $p_{\rm min}$, depends on the number of Higgs fields in the vertex:
\begin{equation}
    p_{\rm min} = \max[N_f + N_X + N_H - n, 0].
    \label{eq:power_v}
\end{equation}
Solving for the power of $E$, we obtain the scaling of the vertex with the maximum power of $E$,
\begin{equation}
    q_{\rm max} = D + d - p_{\rm min}.
    \label{eq:power_E}
\end{equation}
For scaling of sub-leading vertices, additional powers of \( v \) can replace powers of \( E \) while maintaining the total dimensional consistency of the vertex. This substitution arises from the fact that Higgs fields can manifest either as a vev \( v \), which breaks electroweak symmetry, or as a physical Higgs boson. The power of \( v \) for a subleading vertex insertion of the same operator, \( p_k \), and the corresponding power of \( E \), given by \( q_k \), are related to the leading values \( p_{\rm min} \) and \( q_{\rm max} \) as:
\begin{equation}
    p_k = p_{\rm min} + k, \quad q_k = q_{\rm max} - k,
    \label{eq:subleading_pq}
\end{equation}
where \( k \in \lbrace0,1,\ldots,\min[k_{\rm max},q_{\rm max}]\rbrace \) such that $k_{\text{max}}$ is only non-zero when $N_H>0$ and is given by
\begin{equation}
k_{\text{max}} = N_H - \max\left[n - N_f - 2N_X - N_D,\; 0\right]
\end{equation}
The energy scaling of a vertex with $n$-legs of an operator, including subleading terms, can thus be written as:
\begin{equation}
    \mathcal{V}_k^{(n)} \sim \frac{E^{q_k} v^{p_k}}{\Lambda^D},
\end{equation}
where the sum of \( p_k \) and \( q_k \) satisfies $p_k + q_k = D + d$. For \( k = 0 \), the vertex corresponds to the leading contribution, maximizing the energy scaling \( E^{q_{\rm max}} \) and minimizing the dependence on \( v \). For \( k > 0 \), subleading contributions systematically replace powers of \( E \) with \( v \), reducing the energy enhancement but preserving the dimensional consistency of the vertex. We can now define $\lambda$ in the regime $\Lambda\gg E\gg v$ using Eq.~\eqref{eq:scaling_conditions} to be,
\begin{equation}
    \lambda_j^{(n)}=\lambda^{3D-2q_{\rm max}-p_{\rm min}}
\end{equation}
where $(n)$ indicates the number of legs and $j$ labels the SMEFT operator it multiplies and $D$ is the mass dimension of said operator. This dual expansion in $v/\Lambda$ and $E/\Lambda$ allows for a controlled separation of operator contributions based on their energy scaling, complementing the traditional mass dimension hierarchy. The above algorithm can be systematically applied to determine the enhancement for an operator in any tree-level vertex.~\footnote{For the case where the number of fields in the operator matches the number of vertex legs, much of the energy counting can be read off from the spinor-helicity/on-shell expressions~\cite{Shadmi:2018xan, Ma:2019gtx, Aoude:2019tzn, Durieux:2020gip, AccettulliHuber:2021uoa} for SMEFT operators. }

We note here that for vertices with four or more external legs, the kinematics provides independent momentum invariants that are not completely fixed by momentum conservation and on‐shell conditions. This means that when you distribute derivatives across the external legs, each derivative can contribute an independent factor that scales with the energy \(E\). In such cases, the \(E/\Lambda\) expansion reliably captures the energy dependence of the operator contributions.

In contrast, for vertices with $n<4$ the situation is fundamentally different. With only two or three external legs, momentum conservation and on‐shell conditions fully constrain the kinematics, leaving no free invariant to carry an independent factor of \(E\). In these cases, derivative factors are either fully fixed by the particle masses or can be reshuffled using equations of motion and integration‐by‐parts, so the naive \(E/\Lambda\) counting does not necessarily reflect the actual energy scaling. Thus, our counting method is only robust for \(n \ge 4\) because the extra independent kinematic degrees of freedom guarantee that energy enhancements appear as expected, whereas for \(n < 4\) the tight kinematic constraints limit the validity of an \(E/\Lambda\) expansion. For $2$‑ and $3$‑point vertices, we explicitly correct for these limitations, ensuring that the energy scaling tables (see Tables~\ref{tab:2pt_scaling_compact} and \ref{tab:3pt_scaling}) can be used without concern.

\begin{table}[tb]
    \begin{tabular}{|c|c|c|}
        \hline
        Vertex Diagram & Operator & Count \\
        \hline
        \multirow{5}{*}{
            \begin{tikzpicture}
                \filldraw (0,0) circle (2pt);
                \draw (-0.5,0) -- (0,0);
                \draw (0.5,0) -- (0,0);
            \end{tikzpicture}
        } 
         & \( H^8 \) & 1 \\
         & \( H^6 D^2 \) & 2 \\
         & \( H^4 X^2 \) & 10 \\
         & \( \psi^2 H^5 \) & 6 \\
         & \textbf{Total} & \textbf{19} \\
        \hline
        \multirow{5}{*}{
            \begin{tikzpicture}
                \filldraw (0,0) circle (2pt);
                \draw (0,0) -- (0.5,0.5);
                \draw (0,0) -- (0.5,-0.5);
                \draw (0,0) -- (-0.6,0);
            \end{tikzpicture}
        } 
         & \( H^2 X^3 \) & 6 \\
         & \( X H^4 D^2  \) & 6 \\
         & \( \psi^2 H^3 X \) & 22 \\
         & \( \psi^2 H^4 D \) & 13 \\
         & \textbf{Total} & \textbf{47} \\
        \hline
        \multirow{10}{*}{
            \begin{tikzpicture}
                \filldraw (0,0) circle (2pt);
                \draw (0,0) -- (0.5,0.5);
                \draw (0,0) -- (0.5,-0.5);
                \draw (0,0) -- (-0.5,0.5);
                \draw (0,0) -- (-0.5,-0.5);
            \end{tikzpicture}
        } 
         & \( H^4 D^4 \) & 3 \\
         & \( X^4 \) & 43 \\
         & \(  H^2 X^2 D^2 \) & 18 \\
         & \( \psi^2 H X^2  \) & 96 \\
         & \( \psi^2 H^2 D^3 \) & 16 \\
         & \( \psi^2 X^2 D \) & 57 \\
         & \( \psi^2 H^2 X D  \) & 92 \\
         & \( \psi^2 H X D^2 \) & 48 \\
         & \( \psi^2 H^3 D^2 \) & 36 \\
         & \( \psi^4 H^2 \) & 96 \\
         & \( \psi^4 H D \) & 168 \\
         & \( \psi^4 D^2 \) & 67  \\
         & \textbf{Total} & \textbf{740} \\
        \hline
        \multirow{4}{*}{ 
            \begin{tikzpicture}
                \filldraw (0,0) circle (2pt);
                \draw (0,0) -- (0.5,0.5);
                \draw (0,0) -- (0.5,-0.5);
                \draw (0,0) -- (-0.5,0.5);
                \draw (0,0) -- (-0.5,-0.5);
                \draw (0.7,0) -- (0,0);
            \end{tikzpicture}
        } 
        &  &  \\
        & \( \psi^4 X \) & 224 \\
        & \textbf{Total} & \textbf{224} \\
        &  &  \\
        \hline
        Together & \textbf{Total} & \textbf{1030} \\
        \hline
    \end{tabular}
    \caption{Summary of dimension 8 operators and when they start contributing at the vertex level. While the geoSMEFT framework fully re-sums operators contributing to two- and three-point vertices, this resummation does not capture the energy scaling for higher-point vertices. For higher point vertices geoSMEFT re-sums only a small portion which happen not to be energy-enhanced. Our classification explicitly identifies the energy–enhanced operators in these higher-point interactions.}
    \label{fig:geo_org}
\end{table}

\section{Further Enhancements}
\label{sec:further_enhancements}

Our starting point is the baseline assumption that all Wilson coefficients, \(c_j^{(i)}/\Lambda^{i}\), are roughly equal—this reflects a weakly coupled UV theory. However, it can be refined further if desired by considering the UV origin of an operator: operators generated solely at loop level are typically suppressed by an extra factor of order \(1/(16\pi^2)\)~\cite{Arzt_1995, Buchalla:2022vjp,Craig:2019wmo}. Prior literature on tree vs. loop coefficients only covers dimension six and eight. Our tables therefore carry the same assumptions as Ref.~\cite{Craig:2019wmo}, such as neglecting couplings between three heavy fields and a single light field. To categorize operators at dimension ten, one can extend the technique of Ref.~\cite{Craig:2019wmo} with the operator classes generated via Ref.~\cite{Henning:2015alf}. We refrain, however, from applying a naive extension of the tree vs. loop level coefficients to dimension ten in this paper. Thus, our tables from here on out will not include any tree-loop labeling information about dimension ten.

Second, the chirality structure of an operator is important; operators with mixed chirality (e.g. \(e_c L,\, u_c Q\)) tend to be suppressed if flavor universality or minimal flavor violation is assumed~\cite{DAmbrosio:2002vsn, Brivio:2017vri}. To clearly convey these aspects in our summary tables, we have adopted a coding scheme:
\begin{itemize}
    \item Operators with a tree-level origin are typeset plainly.
    \item Operators generated only at loop level are typeset in bold.
    \item Operators with mixed chirality are underlined.
    \item Operators with both features—e.g., loop-induced and mixed chirality—are bold and underlined.
\end{itemize}
We employ this coding throughout the remaining sections. We further note that the $\lambda$ weights are helicity-blind since no assumption is made about whether the vertex leg ends up external or internal; once external $V_L$ or $V_T$ polarisations are fixed for a specific process, one should retain the operator structures that couple to those helicities. For a specific process example, see the energy-enhanced VBF analysis of Ref.\,\cite{Assi:2024zap}.

\section{Energy Scaling Tables}
\label{sec:energy_scaling_tables}

Below we summarize the energy–scaling behavior for SMEFT operators contributing to vertices with varying numbers of external legs. We also incorporate the coding scheme introduced in Section~\ref{sec:further_enhancements}.

Table~\ref{tab:2pt_scaling_compact} presents the scaling for 2-point vertices, which are fully encapsulated within the geoSMEFT framework. In this case, the vertices involve only the standard \(v/\Lambda\) suppression, without additional energy factors. Table~\ref{tab:3pt_scaling} lists the scaling for 3-point vertices; here, most operators continue to scale as \(v^2/\Lambda^2\), with any explicit energy dependence entering only for specific operator classes. As the number of external legs increases, the kinematics becomes non-trivial. Table~\ref{tab:4pt_scaling_replaced} collects the leading \(\lambda\)–scaling for 4-point vertices, while Table~\ref{tab:5pt_scaling_replaced} (for 5-point vertices) and the following Table~\ref{tab:6pt_scaling_replaced} (for 6-point vertices) detail the corresponding scaling for higher multiplicities. 

In all cases, the expressions are organized in terms of the power counting parameter \(\lambda\), which captures the interplay between the process energy \(E\), the electroweak scale \(v\), and the cutoff \(\Lambda\). Only the leading powers in \(\lambda\) are displayed here; subleading corrections have been omitted for clarity, but can be derived from our general algorithm in Section~\ref{sec:energy_expansion}. 

\begin{table}[tb]
    \small
    \renewcommand{\arraystretch}{1.3}
    \begin{tabular}{|c|c|c|}
        \hline
        \multicolumn{3}{|c|}{\textbf{Dimension-6 Operators}} \\
        \hline
        \textbf{Operator} & \(\boldsymbol{\lambda^{(2)}}\) & \textbf{Vertices } \\
        \hline
        \(H^6\)
        & \(\lambda^4\)
        & \(h^2: {v^2}\) \\
        \hline
        \(H^4 D^2\)
        & \(\lambda^4\)
        & \((\partial h)^2: {v^2}\) \\
        \hline
        \(\looponly{H^2 X^2}\)
        & \(\lambda^4\)
        & \((\partial V)^2: {v^2}\) \\
        \hline
        \(\mixedchirality{\psi^2 H^3}\)
        & \(\lambda^4\)
        & \(\psi^2: {v^2}\) \\
        \hline
        \multicolumn{3}{|c|}{\textbf{Dimension-8 Operators}} \\
        \hline
        \textbf{Operator} & \(\boldsymbol{\lambda^{(2)}}\) & \textbf{Vertices } \\
        \hline
        \(H^8\)
        & \(\lambda^8\)
        & \(h^2: {v^4}\) \\
        \hline
        \(H^6 D^2\)
        & \(\lambda^8\)
        & \((\partial h)^2: {v^4}\) \\
        \hline
        \(H^4 X^2\)
        & \(\lambda^8\)
        & \(h^2: {v^4}\) \\
        \hline
        \(\mixedchirality{\psi^2 H^5}\)
        & \(\lambda^8\)
        & \(\psi^2: {v^4}\) \\
        \hline
        \multicolumn{3}{|c|}{\textbf{Dimension-10 Operators}} \\
        \hline
        \textbf{Operator} & \(\boldsymbol{\lambda^{(2)}}\) & \textbf{Vertices } \\
        \hline
        \(H^{10}\)
        & \(\lambda^{12}\)
        & \(h^2: {v^6}\) \\
        \hline
        \(H^8 D^2\)
        & \(\lambda^{12}\)
        & \((\partial h)^2: {v^6}\) \\
        \hline
        \(H^6 X^2\)
        & \(\lambda^{12}\)
        & \(h^2: {v^6}\) \\
        \hline
        \(\mixedchirality{\psi^2 H^7}\)
        & \(\lambda^{12}\)
        & \(\psi^2: {v^6}\) \\
        \hline
    \end{tabular}
    \caption{Energy scaling for 2-point vertices entirely encapsulated in geoSMEFT. We note that there are no new $E$ powers, all just dimension six scaling times powers of $v^2/\Lambda^2$}
    \label{tab:2pt_scaling_compact}
\end{table}

\begin{table}[tb]
    \small
    \renewcommand{\arraystretch}{1.5}
    \begin{tabular}{|c|c|c|}
        \hline
        \multicolumn{3}{|c|}{\textbf{Dimension-6 Operators}} \\
        \hline
        \textbf{Operator} & \(\boldsymbol{\lambda^{(3)}}\) & \textbf{Vertices } \\
        \hline
        \(H^6\)
        & \(\lambda^3\)
        & \(h^3: \displaystyle {v^3}\) \\
        \hline
        \(H^4 D^2\)
        & \(\lambda^3\)
        & \(hVV: \displaystyle {v^3}\) \\
        \hline
        \(\looponly{H^2 X^2}\)
        & \(\lambda\)
        & \(hVV: \displaystyle {v\,E^2}\) \\
        \hline
        \(\mixedchirality{\psi^2 H^3}\)
        & \(\lambda^2\)
        & \(h\,\psi^2: \displaystyle {v^2 E}\) \\
        \hline
        \(\looponly{X^3}\)
        & \(1\)
        & \(VVV: \displaystyle {E^3}\) \\
        \hline
        \(\bothfeatures{\psi^2 H X}\)
        & \(\lambda\)
        & \(\psi^2\,V: \displaystyle {v\,E^2}\) \\
        \hline
        \(\psi^2 H^2 D\)
        & \(\lambda^2\)
        & \(\psi^2\,V: \displaystyle {v^2 E}\) \\
        \hline
        \multicolumn{3}{|c|}{\textbf{Dimension-8 Operators}} \\
        \hline
        \textbf{Operator} & \(\boldsymbol{\lambda^{(3)}}\) & \textbf{Vertices } \\
        \hline
        \(H^8\)
        & \(\lambda^7\)
        & \(h^3: \displaystyle {v^5}\) \\
        \hline
        \(H^6 D^2\)
        & \(\lambda^7\)
        & \(hVV: \displaystyle {v^5}\) \\
        \hline
        \(\looponly{H^2 X^3}\)
        & \(\lambda^4\)
        & \(VVV: \displaystyle {v^2 E^3}\) \\
        \hline
        \(H^4 X^2\)
        & \(\lambda^5\)
        & \(hVV: \displaystyle {v^3 E^2}\) \\
        \hline
        \(H^4 X D^2\)
        & \(\lambda^5\)
        & \(hVV: \displaystyle {v^3 E^2}\) \\
        \hline
        \(\mixedchirality{\psi^2 H^3 X}\)
        & \(\lambda^5\)
        & \(\psi^2\,V: \displaystyle {v^3 E^2}\) \\
        \hline
        \(\mixedchirality{\psi^2 H^5}\)
        & \(\lambda^6\)
        & \(h\,\psi^2: \displaystyle {v^4 E}\)  \\
        \hline
        \(\psi^2 H^4 D\)
        & \(\lambda^5\)
        & \((\partial h)\,\psi^2: \displaystyle {v^3 E^2}\) \\
        \hline
        \multicolumn{3}{|c|}{\textbf{Dimension-10 Operators}} \\
        \hline
        \textbf{Operator} & \(\boldsymbol{\lambda^{(3)}}\) & \textbf{Vertices } \\
        \hline
        \(H^{10}\)
        & \(\lambda^{11}\)
        & \(h^3: \displaystyle {v^7}\) \\
        \hline
        \(H^{8} D^2\)
        & \(\lambda^{11}\)
        & \(hVV: \displaystyle {v^7}\) \\
        \hline
        \(\textcolor{black}{H^4 X^3}\)
        & \(\lambda^{8}\)
        & \(VVV: \displaystyle {v^4 E^3}\) \\
        \hline
        \(H^6 X^2\)
        & \(\lambda^9\)
        & \(hVV: \displaystyle {v^5 E^2}\) \\
        \hline
        \({H^6 X D^2}\)
        & \(\lambda^9\)
        & \(hVV: \displaystyle {v^5 E^2}\) \\
        \hline        
        \(\mixedchirality{\psi^2 H^5 X}\)
        & \(\lambda^9\)
        & \(\psi^2\,V: \displaystyle {v^5 E^2}\) \\
        \hline
        \(\mixedchirality{\psi^2 H^7}\)
        & \(\lambda^{10}\)
        & \(h\,\psi^2: \displaystyle {v^6 E}\) \\
        \hline
        \(\textcolor{black}{\psi^2 H^6 D}\)
        & \(\lambda^{9}\)
        & \((\partial h)\,\psi^2: \displaystyle {v^5 E^2}\) \\
        \hline
    \end{tabular}
    \caption{Energy scaling for 3-point vertices entirely encapsulated in geoSMEFT.  Moving from lower to higher dimension (top to bottom), note that the energy scaling for a given vertex is fixed, so higher dimensional operators only introduce additional powers of $v$.}
    \label{tab:3pt_scaling}
\end{table}

\begin{table*}[tb]
    \centering
    \small
    \renewcommand{\arraystretch}{1.3}
    \begin{tabular}{|c|c|c|}
        \hline
        \multicolumn{3}{|c|}{\textbf{Dimension-6 Operators}} \\
        \hline
        \textbf{Operator} & \(\boldsymbol{\lambda^{(4)}}\) & \textbf{Vertices } \\
        \hline
        \(H^4 D^2\)
        & \(\lambda^2\)
        & \(\displaystyle
        (\partial h)^2 h^2: {E^2},\;\;
        h^3 V: {v E},\;\;
        h^2 V^2: {v^2}
        \) \\
        \hline
        \(\looponly{H^2 X^2}\)
        & \(\lambda^2\)
        & \(\displaystyle
        h^2 V^2: {E^2},\;\;
        h V^3: {v E},\;\;
        V^4: {v^2}
        \) \\
        \hline
        \(\looponly{X^3}\)
        & \(\lambda^2\)
        & \(\displaystyle
        V^4: {E^2}
        \) \\
        \hline
        \(\bothfeatures{\psi^2 H X}\)
        & \(\lambda^2\)
        & \(\displaystyle
        \psi^2 V^2: {v E},\;\;
        \psi^2\,h\,\partial V: {E^2}
        \) \\
        \hline
        \(\psi^2 H^2 D\)
        & \(\lambda^2\)
        & \(\displaystyle
        \psi^2\,V\,h: {v E},\;\;
        \psi^2\,h\,\partial h: {E^2}
        \) \\
        \hline
        \(\psi^4\)
        & \(\lambda^2\)
        & \(\displaystyle
        \psi^4: {E^2}
        \) \\
        \hline
        \multicolumn{3}{|c|}{\textbf{Dimension-8 Operators}} \\
        \hline
        \textbf{Operator} & \(\boldsymbol{\lambda^{(4)}}\) & \textbf{Vertices } \\
        \hline
        \(\looponly{X^4}\)
        & \(\lambda^4\)
        & \(\displaystyle
        V^4: {E^4}
        \) \\
        \hline
        \(H^4 D^4\)
        & \(\lambda^4\)
        & \(\displaystyle
        (\partial h)^4: {E^4},\;\;
        h^3 V: {E^3 v},\;\;
        h^2 V^2: {E^2 v^2},\;\;
        h V^3: {E\,v^3}\;\;
        \) \\
        \hline
        \(\looponly{H^2 X^2 D^2}\)
        & \(\lambda^4\)
        & \(\displaystyle
        (\partial h)^2V^2: {E^4},\;\;
        (\partial h)V^3: {E^3 v},\;\;
        V^4: {E^2 v^2}\;\;
        \) \\
        \hline
        \(\psi^2 H^2 D^3\)
        & \(\lambda^4\)
        & \(\displaystyle
        \psi^2 (\partial h)^2 : {E^4},\;\;
        \psi^2 (\partial h) V: {E^3 v},\;\;
        \psi^2 V^2: {E^2\,v^2}\;\;
        \) \\
        \hline
        \(\looponly{\psi^2 X^2 D}\)
        & \(\lambda^4\)
        & \(\displaystyle
        \psi^2 (\partial^2 V)(\partial V): {E^4}
        \) \\
        \hline
        \(\bothfeatures{\psi^2 H X D^2}\)
        & \(\lambda^4\)
        & \(\displaystyle
        \psi^2 (\partial h)(\partial V): {E^4},\;\;
         \psi^2 (\partial V)^2 : {E^3v}\;\;
        \) \\
        \hline
        \(\psi^4 D^2\)
        & \(\lambda^4\)
        & \(\displaystyle
         \psi^2 (\partial \psi)^2 : {E^4}\;\;
        \) \\
        \hline
        \multicolumn{3}{|c|}{\textbf{Dimension-10 Operators}} \\
        \hline
        \textbf{Operator} & \(\boldsymbol{\lambda^{(4)}}\) & \textbf{Vertices } \\
        \hline
        \(H^4 D^6\)
        & \(\lambda^6\)
        & \(\displaystyle
        \partial^2(\partial h)^4 : {E^6},\;\;
        (\partial h)^3 \partial^2V: {E^5 v},\;\;
        (\partial h)^2 (\partial V)V: {E^4 v^2},\;\;
        (\partial^3 h) V^3: {E^3\,v^3},\;\;
        \partial^2V^4: {E^4\,v^2}\;\;
        \) \\
        \hline
        \(\textcolor{black}{X^4 D^2}\)
        & \(\lambda^6\)
        & \(\partial^2(\partial V)^4 : {E^6}\;\;\) \\
        \hline
        \(\mixedchirality{\psi^2HXD^4}\)
        & \(\lambda^6\)
        & \(\displaystyle
        \partial^2\psi^2(\partial^2 h)(\partial V): {E^6},\;\;
        \partial^2\psi^2(\partial V)^2: {E^5 v}\;\;
        \) \\
        \hline
        \(\textcolor{black}{\psi^2H^2  D^5}\)
        & \(\lambda^6\)
        & \(\displaystyle
        \partial(\partial h)^2(\partial \psi)^2: {E^6},\;\;
        (\partial h)(\partial \psi)^2\partial V: {E^5 v},\;\;
        (\partial \psi)^2\partial V^2: {E^4 v^2}\;\;
        \) \\
        \hline
        \(\psi^4 D^4\)
        & \(\lambda^6\)
        & \(\displaystyle
        (\partial\psi)^4: {E^6}\;\;
        \) \\
        \hline
    \end{tabular}
    \caption{ Energy scaling for 4-point vertices. At a given mass dimension, we only show the operator classes that generate vertices with the highest (leading) power of $\lambda$. Unlike in the previous table, the energy dependence for 4-point vertices can increase as we consider higher and higher dimensional operators. For example, the $V^4$ vertex $\sim E^2$ at dimension six, but grows to $\sim E^4$ at dimension eight and $\sim E^6$ at dimension ten.}
    \label{tab:4pt_scaling_replaced}
\end{table*}

\begin{table*}[tb]
    \centering
    \small
    \renewcommand{\arraystretch}{1.3}
    \begin{tabular}{|c|c|c|}
        \hline
        \multicolumn{3}{|c|}{\textbf{Dimension-6 Operators}} \\
        \hline
        \textbf{Operator} & \(\boldsymbol{\lambda^{(5)}}\) & \textbf{Vertices } \\
        \hline
               \(H^4 D^2\)
        & \(\lambda^4\)
        & \(\displaystyle
        (\partial h)^2 h^2 V: {E},\;\;
        h^3 V^2: {v}\;\;
        \) \\
        \hline
        \(\mixedchirality{\psi^2 H^3}\)
        & \(\lambda^4\)
        & \(\displaystyle
        \psi^2 h^3 : {E}\;\;
        \) \\
        \hline
        \(\looponly{H^2 X^2}\)
        & \(\lambda^4\)
        & \(\displaystyle
        h^2 (\partial V)V^2: {E},\;\;
        h V^4: {v}\;\;
        \) \\
        \hline
        \(\looponly{X^3}\)
        & \(\lambda^4\)
        & \(\displaystyle
        (\partial V)V^4: {E}
        \) \\
        \hline
        \(\bothfeatures{\psi^2 H X}\)
        & \(\lambda^4\)
        & \(\displaystyle
        \psi^2 h V^2: {E}\;\;
        \) \\
        \hline
        \(\psi^2 H^2 D\)
        & \(\lambda^4\)
        & \(\displaystyle
        \psi^2 h^2 V: {E}\; 
        \) \\
        \hline
        \multicolumn{3}{|c|}{\textbf{Dimension-8 Operators} } \\
        \hline
        \textbf{Operator} & \(\boldsymbol{\lambda^{(5)}}\) & \textbf{Vertices } \\
        \hline
        \(\looponly{X^4}\)
        & \(\lambda^6\)
        & \(\displaystyle
        (\partial V)^3V^2: {E^3}
        \) \\
        \hline
        \(H^4 D^4\)
        & \(\lambda^6\)
        & \(\displaystyle
        (\partial h)^3 h V: {E^3},\;\;
        (\partial h)^2 h V^3: {E^2 v},\;\;
        (\partial h) h V^3: {E v^2},\;\;
        h V^4: {v^3}\;\;
        \) \\
        \hline
        \(\looponly{H^2 X^3}\)
        & \(\lambda^6\)
        & \(\displaystyle
        h^2 (\partial V)^3: {E^3},\;\;
         h (\partial V)^2 V^2: {E^2 v},\;\;
         (\partial V) V^4: {E v^2}\;\;
        \) \\
        \hline
        \(\looponly{ H^2 X^2 D^2}\)
        & \(\lambda^6\)
        & \(\displaystyle
        (\partial h)^2(\partial V)V^2: {E^3},\;\;
        (\partial h)(\partial V)^2V^2: {E^2 v},\;\;
        (\partial V)^3V^2: {E v^2}\;\;
        \) \\
        \hline
        \(H^4 X D^2\)
        & \(\lambda^6\)
        & \(\displaystyle
        (\partial h)^2(\partial V)h^2: {E^3},\;\;
        (\partial h)^2hV^2: {E^2 v},\;\;
        (\partial h)hV^3: {E v^2}\;\;
        (\partial h)V^4: {v^3}\;\;
        \) \\
        \hline
        \(\bothfeatures{\psi^2 H X^2 }\)
        & \(\lambda^6\)
        & \(\displaystyle
        \psi^2 h (\partial V)^2 : {E^3},\ ;\;
        \psi^2 (\partial V) V^2: {E^2 v}\;\;
        \) \\
        \hline
        \(\psi^2 H^2 D^3\)
        & \(\lambda^6\)
        & \(\displaystyle
        \psi^2 (\partial h)^2 V : {E^3},\;\;
        \psi^2 (\partial h) V^2: {E^2 v},\;\;
        \psi^2 V^3: {Ev^2}
        \) \\
        \hline
        \(\psi^2 H^2XD\)
        & \(\lambda^6\)
        & \(\displaystyle
        \psi^2 (\partial h)(\partial V)h: {E^3},\;\;
        \psi^2 (\partial h)V^2: {E^2 v},\;\;
        \psi^2 V^3: {E v^2}\;\;
        \)
        \\
        \hline
        \(\looponly{\psi^2 X^2 D}\)
        & \(\lambda^6\)
        & \(\displaystyle
        \psi^2 (\partial V)^2V: {E^3}
        \) 
        \\
        \hline
        \(\bothfeatures{\psi^2 H X D^2}\)
        & \(\lambda^6\)
        & \(\displaystyle
        \psi^2 (\partial h)(\partial V): {E^3},\;\;
         \psi^2 (\partial V)^2 : {E^2v}\;\;
        \) \\
        \hline
        \(\mixedchirality{\psi^2 H^3 D^2}\)
        & \(\lambda^6\) 
        & \(\displaystyle
        \psi^2 (\partial h)^3: {E^3},\;\;
         \psi^2 (\partial h)^2V : {E^2v},\;\;
         \psi^2 (\partial h)V^2 : {Ev^2}\;\;
        \) \\
        \hline
        \(\psi^4 D^2\)
        & \(\lambda^6\)
        & \(\displaystyle
         \psi^3 (\partial \psi) V : {E^3}\;\;
        \) \\
        \hline
        \(\psi^4 X\)
        & \(\lambda^6\)
        & \(\displaystyle
         \psi^4 (\partial V) : {E^3}\;\;
        \) \\
        \hline
        \(\mixedchirality{\psi^4 HD}\) 
        & \(\lambda^6\)
        & \(\displaystyle
         \psi^4 (\partial h) : {E^3},\;\;
         \psi^4 V : {E^2v}\;\;
        \) \\
        \hline
        \multicolumn{3}{|c|}{\textbf{Dimension-10 Operators}} \\
        \hline
        \textbf{Operator} & \(\boldsymbol{\lambda^{(5)}}\) & \textbf{Vertices } \\
        \hline
        \(H^4 D^6\)
        & \(\lambda^8\)
        & \(\displaystyle
        (\partial h)^4 (\partial V): {E^5},\;\;
        (\partial h)^3 (\partial V) V: {E^4 v},\;\;
        (\partial h)^2 (\partial V) V^2: {E^3 v^2},\;\;
        (\partial h) V^5: {E^2\,v^3},\;\;
         (\partial V)V^4: {E\,v^2}\;\;
        \) \\
        \hline
        \({H^4 X D^4}\)
        & \(\lambda^8\)
        & \(\displaystyle
        (\partial h)^4 (\partial V): {E^5},\;\;
        (\partial h)^3 (\partial V)^2: {E^4 v},\;\;
        (\partial h)^2 (\partial V) V^2: {E^3 v^2},\;\;
        (\partial h) (\partial V) V^3: {E^2\,v^3},\;\;
         (\partial V)V^4: {E\,v^2}\;\;
        \) \\
        \hline
        \(\textcolor{black}{H^2 X^3 D^2}\)
        & \(\lambda^8\)
        & \(\displaystyle
        (\partial h)^2 (\partial^2 V)^2(\partial V): {E^5},\;\;
        (\partial h) (\partial^2 V)(\partial V)V: {E^4 v},\;\;
        (\partial V)^3 V^2: {E^3 v^2}\;\;
        \) \\
        \hline
        \(\textcolor{black}{X^5}\)
        & \(\lambda^8\)
        & \(\displaystyle
        (\partial V)^5: {E^5}
        \) \\
        \hline
        \(\textcolor{black}{X^4 D^2}\)
        & \(\lambda^8\)
        & \(\displaystyle
        (\partial V)^5: {E^5}
        \) \\
        \hline
        \(\mixedchirality{\psi^2HXD^4}\)
        & \(\lambda^8\)
        & \(\displaystyle
        \psi^2(\partial^2 h)(\partial V)(\partial V): {E^5}, \;\;    
        \psi^2(\partial V)^3: {E^4v}
        \) \\
        \hline
        \(\mixedchirality{\psi^2H^3D^4}\)
        & \(\lambda^8\)
        & \(\displaystyle
        \psi(\partial\psi)(\partial h)^3: {E^5}, \;\;       
        \psi(\partial\psi)(\partial h)^2V: {E^4v}, \;\;       
        \psi(\partial\psi)(\partial h)V^2: {E^3v^2}, \;\;       
        \psi(\partial\psi)V^3: {E^2v^3}
        \) \\
        \hline
        \(\mixedchirality{\psi^2HX^2D^2}\)
        & \(\lambda^8\)
        & \(\displaystyle
        \psi^2(\partial h)(\partial^2 V)V: {E^5}, \;\;       
        \psi^2(\partial V)V^2: {E^4v}
        \) \\
        \hline
        \(\textcolor{black}{\psi^2H^2D^5}\)
        & \(\lambda^8\)
        & \(\displaystyle
        \psi^2(\partial^2 h^2)(\partial^2 V): {E^5}, \;\;       
        \psi^2(\partial h)(\partial^2 V)V: {E^4v}, \;\;       
        \psi^2(\partial h)(\partial V)V^2: {E^3v^2}
        \) \\
        \hline
        \({\textcolor{black}{\psi^2H^2XD^3}}\)
        & \(\lambda^8\)
        & \(\displaystyle
        \psi^2(\partial^2 h^2)(\partial^2 V): {E^5}, \;\;       
        \psi^2(\partial^2 h)(\partial V)V: {E^4v}, \;\;       
        \psi^2(\partial h)(\partial V)V^2: {E^3v^2}
        \) \\
        \hline
        \(\textcolor{black}{\psi^2X^3D}\)
        & \(\lambda^8\)
        & \(\displaystyle
        \psi(\partial \psi)(\partial V)^3: {E^5} \;\;       
        \) \\
        \hline
        \(\psi^4D^4\)
        & \(\lambda^8\)
        & \(\displaystyle
         (\psi\partial\psi)(\psi\partial^2\psi) V: {E^5}
        \) \\
        \hline
        \({\psi^4XD^2}\)
        & \(\lambda^8\)
        & \(\displaystyle
      (\psi\partial\psi)(\psi\partial^2\psi)\partial V: {E^5}
        \) \\
        \hline
        \(\mixedchirality{\psi^4 H D^3}\)
        & \(\lambda^8\)
        & \(\displaystyle
         (\psi\partial\psi)(\psi\partial\psi)(\partial h): {E^5}, \;\;       
        (\psi\partial\psi)(\psi\partial\psi)V: {E^4v}
        \) \\
        \hline
    \end{tabular}
    \caption{ Energy scaling for 5-point vertices. We only display operator classes with leading powers in $\lambda$, the remainder are subleading. As in Table~\ref{tab:4pt_scaling_replaced}, the energy dependence for these vertices can change as we vary the operator dimension.}
    \label{tab:5pt_scaling_replaced}
\end{table*}

\begin{table*}[tb]
    \centering
    \small
    \renewcommand{\arraystretch}{1.3}
    \begin{tabular}{|c|c|c|}       
        \hline
        \multicolumn{3}{|c|}{\textbf{Dimension-6 Operators}} \\
        \hline
        \textbf{Operator} & \(\boldsymbol{\lambda^{(6)}}\) & \textbf{Vertices } \\
        \hline
               \(H^4 D^2\)
        & \(\lambda^6\)
        & \(\displaystyle
        h^2 V^4: {1}\;\;
        \) \\
        \hline
        \(\mixedchirality{\psi^2 H^3}\)
        & \(\lambda^6\)
        & \(\displaystyle
        \psi^2 h^3 : {1}\;\;
        \) \\
        \hline
        \(\looponly{H^2 X^2}\)
        & \(\lambda^6\)
        & \(\displaystyle
        h^2 V^4: {1}
        \) \\
        \hline
        \(\looponly{X^3}\)
        & \(\lambda^6\)
        & \(\displaystyle
        V^6: {1}
        \) \\
        \hline
        \multicolumn{3}{|c|}{\textbf{Dimension-8 Operators}} \\
        \hline
        \textbf{Operator} & \(\boldsymbol{\lambda^{(6)}}\) & \textbf{Vertices } \\
        \hline
        \(\looponly{X^4}\)
        & \(\lambda^8\)
        & \(\displaystyle
        (\partial V)^2V^4: {E^2}
        \) \\
        \hline
           \(H^6 D^2\)
        & \(\lambda^8\)
        & \(\displaystyle
        (\partial h)^2 h^4: {E^2},\;\;
        (\partial h) h^4 V: {E v},\;\;
         h^4 V^2: {v^2}\;\;
        \) \\
        \hline
        \(H^4 D^4\)
        & \(\lambda^8\)
        & \(\displaystyle
        (\partial h)^2 h^2 V^2: {E^2},\;\;
        (\partial h)h^3 V^3: {E v},\;\;
         h^2 V^4: {v^2}\;\;
        \) \\
        \hline
        \(\looponly{H^2 X^3}\)
        & \(\lambda^8\)
        & \(\displaystyle
        (\partial V)^2 h^2 V^2: {E^2},\;\;
        (\partial V) h V^4: {E v},\;\;
        V^6: {v^2}\;\;
        \) \\
        \hline
       \(\textcolor{black}{H^4 X^2 }\)
        & \(\lambda^8\)
        & \(\displaystyle
        (\partial V)^2 h^4: {E^2},\;\;
        (\partial V) V^2 h^3: {E v},\;\;
        V^4 h^2: {v^2}\;\;
        \) \\
        \hline
        \(\looponly{H^2 X^2  D^2}\)
        & \(\lambda^8\)
        & \(\displaystyle
        h^2(\partial^2 V)V^3: {E^2},\;\;
        h(\partial V)V^4: {E v},\;\;
        V^6: {v^2}\;\;
        \) \\
        \hline
        \(H^4 X D^2\)
        & \(\lambda^8\)
        & \(\displaystyle
        h^4(\partial V)^2: {E^2},\;\;
        h^3(\partial V) V^2: {E v},\;\;
        h^2V^4:v^2\;\;
        \) \\
        \hline
        \(\bothfeatures{\psi^2 H X^2 }\)
        & \(\lambda^8\)
        & \(\displaystyle
        \psi^2h(\partial V) V^2: {E^2},\;\;
        \psi^2 V^4: {E v}\;\;
        \) 
        \\
        \hline
        \(\looponly{\psi^2 X^2 D}\)
        & \(\lambda^8\)
        & \(\displaystyle
        \psi^2 (\partial V)V^3: {E^2}
        \) 
        \\
        \hline
       \(\mixedchirality{\psi^2 H^3 X}\)
        & \(\lambda^8\)
        & \(\displaystyle
        \psi^2h^3(\partial V) : {E^2},\;\;
        \psi^2h^2 V^2: {E v}\;\;
        \) \\
        \hline
        \(\psi^2 H^2 D^3\)
        & \(\lambda^8\)
        & \(\displaystyle
        \psi^2h(\partial V) V^2: {E^2},\;\;
        \psi^2 V^4: {E v}\;\;
        \) \\
        \hline
       \(\psi^2 H^4 D\)
        & \(\lambda^8\)
        & \(\displaystyle
        \psi^2h^3(\partial h) : {E^2},\;\;
        \psi^2 h^3 V: {E v}\;\;
        \) \\
        \hline
        \(\psi^2 H^2XD\)
        & \(\lambda^8\)
        & \(\displaystyle
        \psi^2h^2(\partial V)V : {E^2},\;\;
        \psi^2 h V^3: {E v}\;\;
        \) \\
        \hline
        \(\bothfeatures{\psi^2 H X D^2}\)
        & \(\lambda^8\)
        & \(\displaystyle
        \psi^2h(\partial V)V^3 : {E^2},\;\;
        \psi^2V^4: {E v}\;\;
        \) \\
        \hline
        \(\mixedchirality{\psi^2 H^3 D^2}\)
        & \(\lambda^8\)
        & \(\displaystyle
        \psi^2h^3(\partial V) : {E^2},\;\;
        \psi^2h^2V^2 : {E v}\;\;
        \) \\
        \hline
       \(\psi^4 H^2\)
        & \(\lambda^8\)
        & \(\displaystyle
         \psi^4h^2 : {E^2}\;\;
        \) \\
        \hline
        \(\psi^4 D^2\)
        & \(\lambda^8\)
        & \(\displaystyle
         \psi^4 V^2 : {E^2}\;\;
        \) \\
        \hline
        \(\psi^4 X\)
        & \(\lambda^8\)
        & \(\displaystyle
         \psi^4 V^2 : {E^2}\;\;
        \) \\
        \hline
        \(\mixedchirality{\psi^4 HD}\)
        & \(\lambda^8\)
        & \(\displaystyle
         \psi^4 h V : {E^2}\;\;
        \) \\
        \hline
    \end{tabular}
    \caption{ Energy scaling for 6-point vertices. We only show the operator classes with leading powers in $\lambda$, and up to dimension 8 in this case since the leading 6-point dimension 10 scaling is highly subleading in $\lambda^{10}$. As in Table~\ref{tab:4pt_scaling_replaced}, the energy scaling of a particular vertex can increase as we consider operators of higher and higher dimension.}
    \label{tab:6pt_scaling_replaced}
\end{table*}

\section{Example processes}
\label{sec:examples}

With our dual expansion and power–counting framework in hand, we now turn to processes with four, five, and six external legs—scenarios that are particularly relevant for HL–LHC studies~\cite{Biekotter:2020flu,Bishara:2022vsc,Celada:2024mcf}. In these cases, the richer kinematics allow independent energy scales to appear. Moreover, aspects such as chirality, gauge boson polarization, and interference with the SM only become evident when you analyze a specific process. For a given process, we are interested in the scaling of different (SM and SMEFT)  contributions to the amplitude. To determine the scaling, we collect coupling and $\lambda$ factors from each diagram/topology that can participate. We will ignore external states (spinors,polarizations, etc.) which are common to all contributions to a given amplitude and result in an overall kinematic factor which can be dimensionful.

\subsection{$pp\rightarrow VH(q\bar{q})$}\label{sec:VH}

In this subsection, we apply the $\lambda$ power counting to associated to 4-point associated Higgs production. At tree level in the SM and assuming a $q\bar{q}$ initial state, the process $pp\to VH$ proceeds via a single diagram: a $\bar{q}qV$ vertex is connected to a $hVV$ vertex by a vector boson propagator. However, beyond the SM, one must also consider additional topologies that involve a single four-particle vertex, namely a $\bar{q}qVh$ interaction.

For each topology, the vertices can be expanded in a series in $\lambda$ (with $\lambda\ll 1$). For example, the $\bar{q}qV$ on-shell vertex  scales as
\begin{align}
\label{eq:vhtrips}
g_{\bar{q}qV} &= \frac{g^{\rm SM}_{\bar{q}qV}}{\lambda^2} 
+ \frac{\lambda}{\hat\Lambda^2}\, \bothfeatures{c_{\psi^2HX}}
+ \frac{\lambda^2}{\hat\Lambda^2}\, c_{\psi^2H^2D} \nonumber\\[1mm]
&\quad + \frac{\lambda^5}{\hat\Lambda^4}\, \bothfeatures{c_{\psi^2H^3X}}
+ \frac{\lambda^5}{\hat\Lambda^4}\, c_{\psi^2H^4D} + \cdots \,.
\end{align}
where we factor out the common dimensionful kinematic factor in order to focus on the scaling. Here the factor \(1/\lambda^{2}\) originates from the two external on-shell spinor's wave-function normalization, each supplying \(\sqrt{E}\sim\lambda^{-1}\) in our power counting. The coefficients $c_{\mathcal O}$ (with $\mathcal O$ labeling the operator from Tables~\ref{tab:3pt_scaling},\ref{tab:4pt_scaling_replaced}) indicate the contribution of the corresponding operator (including the coding from Sect.~\ref{sec:further_enhancements}). The factors of $\hat\Lambda$ serve as bookkeeping devices to keep track of the order of the various operator contributions, i.e. re-writing $c^{(i)}_j$ in Eq.~\eqref{eq:SMEFT_n_point} to $c_j/{\hat\Lambda}^i$. In Eq.~\eqref{eq:vhtrips}, only the first few orders in $\lambda$ are shown.

\begin{figure}[tb]
    \centering
    \includegraphics[width=0.85\linewidth]{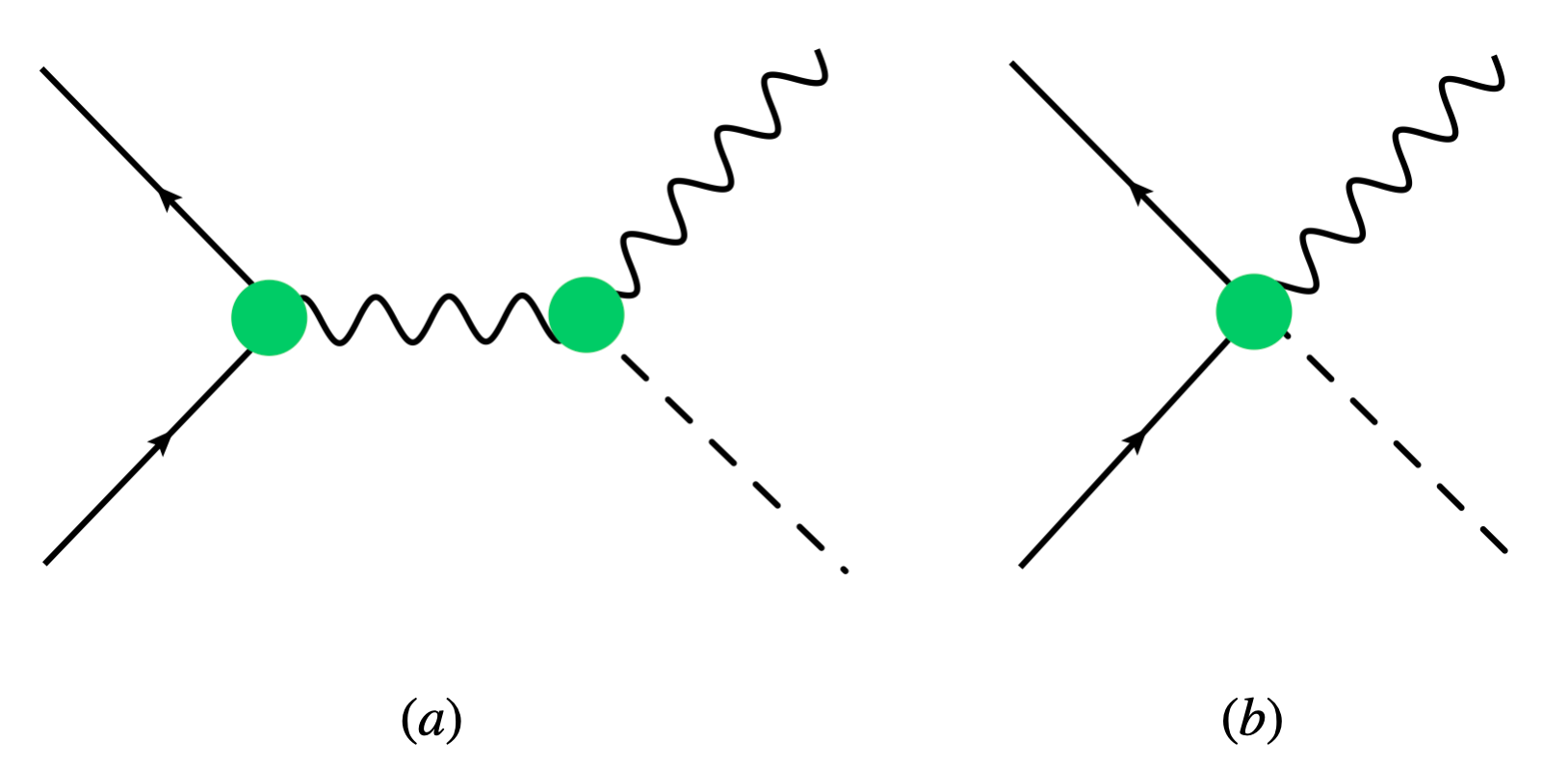}
    \caption{Topologies for $\bar q q \to VH$ production labeled by 3-point $(a)$ and 4-point $(b)$ vertex insertions.}
    \label{fig:VH}
\end{figure}

By similar reasoning, the expansion for the scaling of the $hVV$ vertex is given by
\begin{align}
g_{hVV} &= \frac{g^{\rm SM}_{hVV}}{\lambda} 
+ \frac{\lambda}{\hat\Lambda^2}\, \looponly{c_{H^2X^2}}
+ \frac{\lambda^3}{\hat\Lambda^2}\, c_{H^4D^2} \nonumber\\[1mm]
&\quad + \frac{\lambda^5}{\hat\Lambda^2}\, c_{H^4XD^2}
+ \frac{\lambda^5}{\hat\Lambda^4}\, c_{H^4X^2} + \cdots\,,
\end{align}
while the four-point vertex scaling (which has no SM counterpart) is
\begin{align}
g_{\bar{q}qhV} &= \frac{\lambda^2}{\hat\Lambda^2}\, \bothfeatures{c_{\psi^2HX}}
+ \frac{\lambda^3}{\hat\Lambda^2}\, c_{\psi^2H^2D}
+ \frac{\lambda^4}{\hat\Lambda^4}\, \bothfeatures{c_{\psi^2HXD^2}} \nonumber\\[1mm]
&\quad + \frac{\lambda^5}{\hat\Lambda^4}\, c_{\psi^2H^2D^3}
+ \frac{\lambda^5}{\hat\Lambda^4}\, c_{\psi^2H^4D} + \cdots\,.
\end{align}
Note that the SM contribution to $g_{hVV}$ carries a weight of $1/\lambda$ since it is proportional to $v$ rather than an energy $E$. Moreover, we include sub-leading vertices, such as those that scale as $\lambda^3$ and $\lambda^5$ since in this case they end up providing the leading contributions once combined in the amplitude. These sub-leading vertices are not presented in the tables but can be simply determined from the counting algorithm.

Multiplying the $g_{\bar{q}qV}$ and $g_{hVV}$ expansions and including a factor of $E^{-2}$ which scales as $\lambda^4$ from the vector boson propagator, the net scaling for the topology on the left in Fig.~\ref{fig:VH} becomes (to $\mathcal{O}(\lambda^5)$)
\begin{align}
\mathcal{A}^{(1)}_{qqVh} &= \lambda\, g^{\rm SM}_{\bar{q}qV}\, g^{\rm SM}_{hVV}
+ \frac{\lambda^3}{\hat\Lambda^2}\, g^{\rm SM}_{\bar{q}qV}\, \looponly{c_{H^2X^2}}
+ \frac{\lambda^4}{\hat\Lambda^2}\, g^{\rm SM}_{hVV}\, \bothfeatures{c_{\psi^2HX}} \nonumber\\[1mm]
&\quad + \frac{\lambda^5}{\hat\Lambda^2}\,\Bigl(g^{\rm SM}_{hVV}\, c_{\psi^2H^2D}
+ g^{\rm SM}_{\bar{q}qV}\, c_{H^4D^2}\Bigr) + \cdots\,.
\end{align}
For the second topology, the scaling is given directly by
\[
\mathcal{A}^{(2)}_{qqVh} = g_{\bar{q}qhV}\,.
\]

A comparison of $\mathcal{A}^{(1)}_{qqVh}$ and $\mathcal{A}^{(2)}_{qqVh}$, by counting powers of $\lambda$, indicates that the largest SMEFT effect (i.e. the contribution with the fewest powers of $\lambda$) arises from the right hand topology in Fig~\ref{fig:VH}. However, several refinements are possible. For example, operators whose interference with the SM is suppressed—either because they have a different fermion chirality structure or because they contribute predominantly to transversely polarized vector bosons (in contrast to the predominantly longitudinally polarized $V$ in the SM)—can be neglected. This prescription eliminates the operators shaded in purple, as well as $\looponly{c_{H^2X^2}}$ and $c_{H^4X^2}$. 

\begin{figure*}[htp!]
    \centering
    \includegraphics[width=0.9\linewidth]{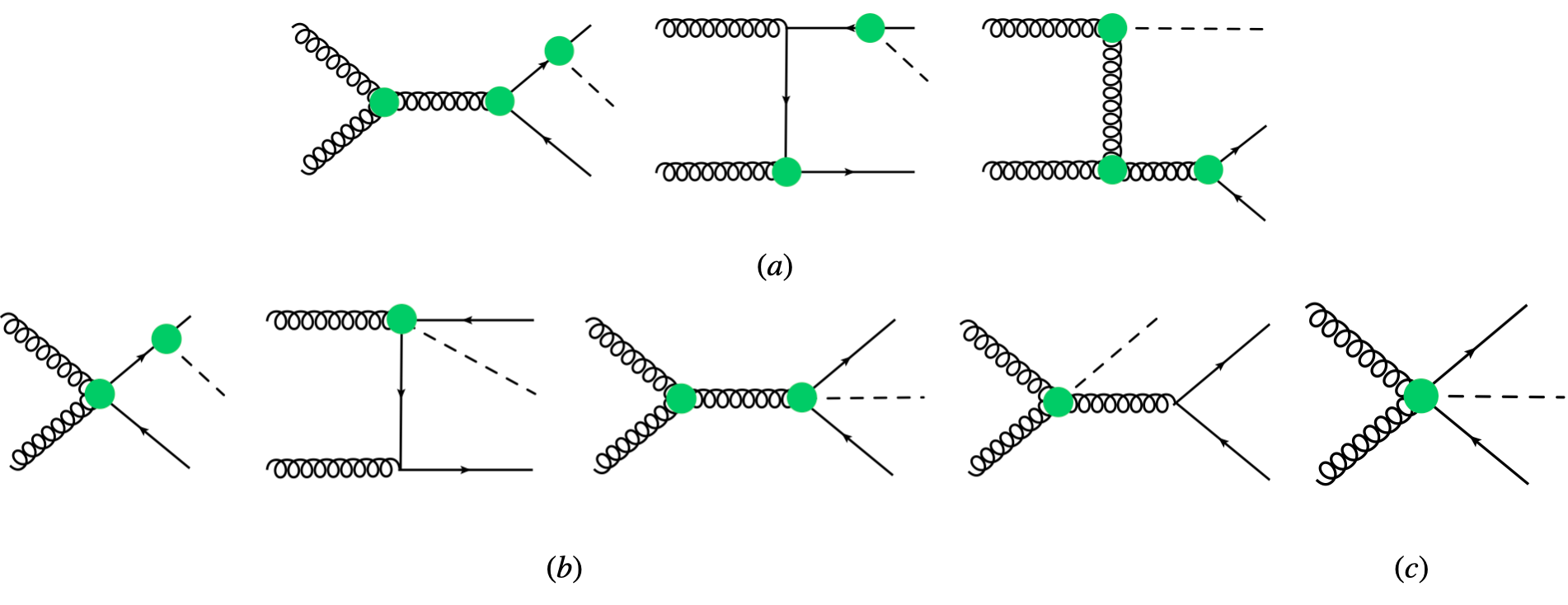}
    \caption{Topologies for $gg\to t\bar{t}H$ labeled by only 3-point $(a)$, mixed 3-point and 4-point $(b)$, and 5-point $(c)$ vertex insertions.}
    \label{fig:ttH}
\end{figure*}

Two of the remaining operator classes, \(c_{\psi^2H^2D}\) and \(\looponly{c_{H^2X^2}}\), are tightly constrained by \(Z\)-pole and Higgs pole measurements~\cite{Ellis:2018gqa,Ellis:2020unq,Bartocci:2023nvp,Giani:2023gfq}. One way to comply with these constraints is to assume that the scale $\Lambda$ is so large that $c \sim \mathcal O(1)$ is consistent. This results in a smaller $\lambda$, suppressing all SMEFT effects. However, from the bottom-up perspective, it is also interesting to consider a scenario where \(c_{\psi^2H^2D}\) and \(\looponly{c_{H^2X^2}}\) are small while the other Wilson coefficients remain $\mathcal O(1)$. In this case, higher order operators can rise in importance. 

Under these assumptions, which are consistent with those in Refs.~\cite{Hays:2018zze, Corbett:2023yhk}, the scaling of the amplitudes simplify to
\begin{align}
\mathcal{A}^{(1)}_{qqVh} &= \lambda\, g^{\rm SM}_{\bar{q}qV}\, g^{\rm SM}_{hVV}
+ \frac{\lambda^5}{\hat\Lambda^2}\, g^{\rm SM}_{\bar{q}qV}\, c_{H^4D^2} + \cdots\,, \nonumber\\[1mm]
\mathcal{A}^{(2)}_{qqVh} &= \frac{\lambda^5}{\hat\Lambda^4}\, c_{\psi^2H^2D^3}
+ \frac{\lambda^5}{\hat\Lambda^4}\, c_{\psi^2H^4D} + \cdots\,,
\end{align}
resulting in identical SMEFT scaling for both topologies. Moreover, our analysis underscores the special importance of operators of the type \(\psi^2H^2D^3\); their pronounced energy dependence and distinctive interference patterns make them especially promising probes of new physics, as also emphasized in Refs.~\cite{Corbett:2023yhk,Degrande:2023iob,Assi:2024zap}.

Of course, alternative assumptions could be made to see how the dominant SMEFT contributions (i.e. those with the fewest $\lambda$ powers) might change. In our treatment, we have effectively set the “removed” operators to zero rather than simply suppressing them, but other approaches are possible.

\subsection{$gg\to t\bar t H$}
\label{sec:ttg}

As a second application, we consider the 5-point process $gg \to t\bar{t}h$. This process has been analyzed to some extent within the SMEFT framework but never in a fully systematic manner. Moreover, it involves five particles at the parton level and thus serves as an example of the five-particle vertex scaling discussed in Table~\ref{tab:5pt_scaling_replaced}.

By inspection of Fig.~\ref{fig:ttH} there are several distinct topologies:
\begin{itemize}
    \item Three topologies involve only three-point vertices.
    \item Four topologies involve one four-point vertex in combination with a three-point vertex.
    \item One topology involves a five-particle vertex.
\end{itemize}
As seen in Tables~\ref{tab:3pt_scaling},\ref{tab:4pt_scaling_replaced},\ref{tab:5pt_scaling_replaced} and in the previous example, vertices with fewer particles generally come with smaller intrinsic powers of $\lambda$, but must be connected via one or more propagators (which introduce additional $\lambda$ factors) to build the full amplitude. In contrast, higher-point vertices inherently carry larger powers of $\lambda$.

\begin{figure*}[htp!]
    \centering
    \includegraphics[width=0.9\linewidth]{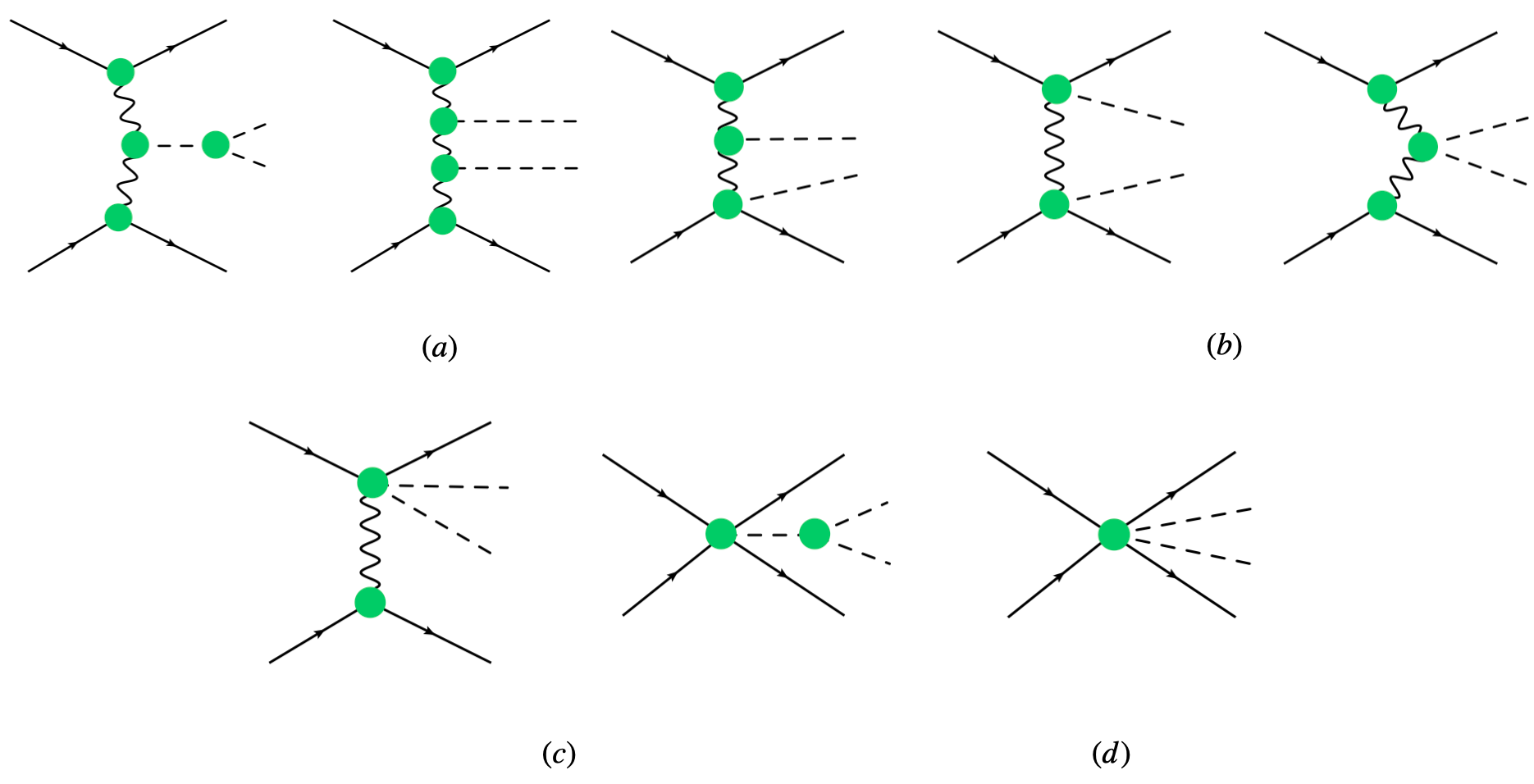}
    \caption{Topologies for di-Higgs production labeled by only 3-point $(a)$, mixed 3-point and 4-point $(b)$, mixed 3-point and 5-point $(c)$, and 6-point $(d)$ vertex insertions.}
    \label{fig:dihiggs}
\end{figure*}

Working to $\mathcal{O}(\lambda^6)$, the different topologies scale as\footnote{Note that all propagators here scale as $E^{-2}$, as the $\lambda$ counting treats $n$-particle vertices as amplitudes which should be glued together to make larger amplitudes~\cite{Elvang:2013cua}.}
\begin{align}
\mathcal{A}^{(1)}_{ggttH} &= \lambda^2\, g^{\rm SM}_{VVV}\, g^{\rm SM}_{\bar q q h}\, g^{\rm SM}_{\bar q q V}  \nonumber\\[1mm]
&\quad
+\frac{\lambda^4}{\hat \Lambda^2}\, g^{\rm SM}_{\bar q q V}\, g^{\rm SM}_{\bar q q h}\, \looponly{c_{X^3}} 
+\frac{\lambda^5}{\hat \Lambda^2}\, g^{\rm SM}_{VVV}\, g^{\rm SM}_{\bar q q h}\, \bothfeatures{c_{\psi^2 H X}} \nonumber\\[1mm]
&\quad + \frac{\lambda^6}{\hat \Lambda^2}\, g^{\rm SM}_{VVV}\,\Bigl( g^{\rm SM}_{\bar q q h}\, c_{\psi^2 H^2 D} +  g^{\rm SM}_{\bar q q V}\,\mixedchirality{c_{\psi^2H^3}}\Bigr)
+\cdots\,, \\
\mathcal{A}^{(2)}_{ggttH} &= \lambda^2\, g^{\rm SM}_{\bar q q h}\,\Bigl(g^{\rm SM}_{\bar q q V}\Bigr)^2  
+\frac{\lambda^5}{\hat \Lambda^2}\, g^{\rm SM}_{\bar q q V}\, g^{\rm SM}_{\bar q q h}\, \bothfeatures{c_{\psi^2 H X}} \nonumber\\[1mm]
&\quad + \frac{\lambda^6}{\hat \Lambda^2}\, g^{\rm SM}_{\bar q q V}\,\Bigl( g^{\rm SM}_{\bar q q h}\, c_{\psi^2 H^2 D} +  g^{\rm SM}_{\bar q q V}\,\mixedchirality{c_{\psi^2H^3}}\Bigr)
+\cdots\,, \\
\mathcal{A}^{(3)}_{ggttH} &=  \frac{\lambda^5}{\hat \Lambda^2}\, g^{\rm SM}_{VVV}\, g^{\rm SM}_{\bar q q V}\, \looponly{c_{H^2X^2}} + \cdots\,, \\
\mathcal{A}^{(4)}_{ggttH} &=  \frac{\lambda^6}{\hat \Lambda^4}\, g^{\rm SM}_{\bar q q h}\, \mixedchirality{c_{\psi^2X^2D}} + \cdots\,, \\
\mathcal{A}^{(5)}_{ggttH} &= \frac{\lambda^4}{\hat \Lambda^2}\, g^{\rm SM}_{\bar q q V}\, \bothfeatures{c_{\psi^2HX}} 
  + \frac{\lambda^6}{\hat \Lambda^4}\, g^{\rm SM}_{\bar q q V}\, \bothfeatures{c_{\psi^2HXD^2}} +\cdots\,, \\
\mathcal{A}^{(6)}_{ggttH} &= \frac{\lambda^4}{\hat \Lambda^2}\, g^{\rm SM}_{VVV}\, \bothfeatures{c_{\psi^2HX}}  
 \nonumber\\[1mm]
&\quad + \frac{\lambda^6}{\hat \Lambda^4}\,\Bigl( g^{\rm SM}_{VVV}\, \bothfeatures{c_{\psi^2HXD^2}} + \looponly{c_{X^3}}\,\bothfeatures{c_{\psi^2HX}}\Bigr) + \cdots\,, \\
\mathcal{A}^{(7)}_{ggttH} &= \frac{\lambda^5}{\hat \Lambda^2}\, g^{\rm SM}_{\bar q q V}\, \looponly{c_{H^2X^2}} + \cdots\,, \\
\mathcal{A}^{(8)}_{ggttH} &= \frac{\lambda^4}{\hat \Lambda^2}\,\bothfeatures{c_{\psi^2HX}}  
+\frac{\lambda^6}{\hat \Lambda^4}\, \bothfeatures{c_{\psi^2 H X^2}}\,.
\end{align}

As in the $pp\to VH$ example, one can explore how different assumptions affect the relative importance of the SMEFT terms. For instance, if one drops the suppressed (or non-interfering) terms, the lowest-order SMEFT contribution originates from the four-point vertex proportional to $\lambda^5\, \looponly{c_{H^2X^2}}$ in the topologies labeled a.) and c.) in Fig.~\ref{fig:ttH}. However, since operators of the $\looponly{c_{H^2X^2}}$ class are constrained by Higgs-pole measurements, setting $\looponly{c_{H^2X^2}}$ to zero shifts the leading SMEFT contributions to $\mathcal{O}(\lambda^6)$, five-particle interaction which arises from the dimension-eight operator $\looponly{c_{\psi^2X^2D}}$.

It is important to emphasize that these assumptions are  illustrative. The treatment of suppressed couplings (by effectively setting them to zero) is meant to simplify the analysis, yet the resulting scaling behavior still points to interesting contributions from operators that have not been included in previous studies. 

Alternatively, if we zero all operators generated at loop level, then the leading SMEFT terms are $\lambda^5\, c_{\psi^2 H^2D}$ and $\lambda^6\, \looponly{c_{\psi^2 H^3}}$. If these are somehow suppressed -- given that $c_{\psi^2 H^2D}$ and $\looponly{c_{\psi^2 H^3}}$ are bounded by $Z$-pole and Higgs pole measurements, respectively -- the next terms in the $\lambda$ expansion (all $\mathcal O(\lambda^7)$) are $c_{H^4D^2}$ at dimension six and $c_{\psi^2H^2D^3}, c_{\psi^2H^2XD}, c_{\psi^2 H^4D}$ at dimension eight.

\subsection{$\bar q q\rightarrow jjHH$}\label{sec:dihiggs}

As a third application, we consider di-Higgs production via vector boson fusion, $\bar q q\rightarrow jjHH$. We choose this process as it (at tree level, lowest order) involves six particles and therefore 6-point vertices. Additionally, it is part of the current (and future) LHC di-Higgs program~\cite{Arganda:2018ftn,Cepeda:2019klc,Chen:2021rid,Gomez-Ambrosio:2022qsi, CMS:2022gjd, Cappati:2022skp, Delgado:2023ynh}, which aims to extract the Higgs self-coupling.

Vector boson fusion processes such as  $\bar q q\rightarrow jjHH$ impose strict cuts on the final state jets, requiring them to be well separated and to have a large $m_{jj}$. These cuts suppress QCD production of $pp \to jj + X$, where $X$ is some combination of massive electroweak bosons. As explored in Ref.~\cite{Banerjee:2019twi}, these cuts retain SMEFT topologies with the SM-like t-channel propagator(s), while suppressing s-channel topologies where the final jet(s) are disconnected from the initial quarks/anti-quarks. To simplify this example, we will omit suppressed ($s$-channel) SMEFT topologies.

 We will also omit topologies involving $g_{\psi^2h}$, as the SM contribution is suppressed by the light quark Yukawas, as well as topologies with $g_{\psi^2h^2}$, which would contribute to $\bar q q \to h h$. The remaining eight topologies are shown in Fig.~\ref{fig:dihiggs} and contain three-particle through six-particle interactions.

 Carrying out the $\lambda$ power counting, we find that the dominant SM contribution (at large energy) is the rightmost topology b.) of Fig~\ref{fig:dihiggs}, scaling as $\lambda^4$. Without additional assumptions, the dominant SMEFT contribution comes at $\mathcal O(\lambda^6)$ from $\looponly{c_{H^2X^2}}$, which modify the $H^2V^2$ coupling. If we zero these operators to take into account the fact that they are bounded from $H \to \gamma\gamma$ measurements, while also zeroing operators either bounded by LEP (such as $c_{\psi^2H^2D}$ ) or with non-SM helicity structure, the first SMEFT effects to enter are $\mathcal O(\lambda^8)$ and come from operator classes $c_{H^4D^2}$ at dimension six and $c_{\psi^2H^2D^3}, c_{\psi^2H^2XD}, c_{\psi^4H^2}$ at dimension eight. The impact of two of these three ($c_{\psi^2H^2D^3}$ and $c_{\psi^4H^2}$) on high-energy single Higgs VBF ($pp \to jjH$) production was recently explored in Ref.~\cite{Assi:2024zap}. These operators were carefully studied in that work and confirmed to be the most significant contributors, thereby validating our prescription.  

\section{Discussion}
\label{sec:discussion}

In this work we have developed an energy–enhanced expansion of the SMEFT that supplements the usual $v/\Lambda$ power counting with an explicit expansion in the energy scale $E/\Lambda$. By introducing a dimensionless parameter $\lambda$ to systematically track the interplay between the electroweak scale $v$, the process energy $E$, and the new physics scale $\Lambda$, we have organized SMEFT operators according to the number of external legs in their corresponding vertices. 
This vertex–based classification is motivated by the energy versus vev counting in 2– and 3–point interactions elucidated by the geometric SMEFT organization scheme, and it neatly formalises and extends this counting to higher-point vertices.


Our approach reveals that operators with fewer powers of $v$ and correspondingly higher powers of $E$ are naturally energy–enhanced, a feature that becomes increasingly important in high–energy processes. In the examples we look at, we demonstrated how different topologies contribute with distinct $\lambda$ scalings, and how—with suitable assumptions about interference and chirality structures—the leading SMEFT effects can be isolated. This analysis clarifies which higher–dimensional operators are most likely to affect the kinematic tails of differential distributions.

By systematically identifying energy–enhanced operators, our framework highlights the contributions that grow with energy and are thus most likely to produce observable deviations in high–energy collider experiments. Models that generate such energy–enhanced operators --  especially those arising at tree level -- can be more readily probed in differential measurements where the energy scale $E$ is large. In this way, our classification not only reduces the effective operator basis that experimentalists need to consider, but also provides theoretical guidance on which signatures should be prioritized in searches for physics beyond the SM.

Several caveats are in order. First, throughout this work we assume that new physics is weakly coupled—that is, all Wilson coefficients \(c_i/\Lambda\) are taken to be of comparable magnitude (apart from the standard loop suppression). Second, our power counting is based on on-shell conditions, so vertices that vanish by the equations of motion are not properly ordered within this scheme. Third, while our analysis is carried out to leading order in \(\lambda\), sub-leading corrections may be relevant for certain observables and warrant further study. Thus, the \(\lambda\) counting we advocate for is a guide; large numerical or coupling prefactors that appear in explicit calculations can override it, so final priorities on what orders in $\lambda$ to keep should be set after computing the full amplitude. Finally, our illustrative assumptions—such as setting to zero those operators whose interference with the SM is suppressed—are chosen to simplify the discussion; a more complete phenomenological analysis would need to account for additional effects from flavor structure, loop-level matching, and off-shell dynamics.

Overall, our energy–enhanced expansion provides a fresh take on organizing SMEFT operators beyond simple mass–dimension counting. By highlighting the operators with the strongest energy growth at tree level, it narrows the effective basis down to the most promising candidates for new physics. This, in turn, motivates targeted experimental analyses that focus on the energy–enhanced observables where deviations from the SM are most likely to show up.

\section*{Acknowledgements}

We thank Kelci Mohrman, Pavel Nadolsky and Nick Smith for their helpful feedback. The work of AM is partially supported by the National Science Foundation under Grant Numbers PHY--2112540 and PHY--2412701. The work of BA is supported by the DOE grant number DE--SC0011784 and NSF grant number OAC--2103889 as well as the Fermi National Accelerator Laboratory (Fermilab). This manuscript has been authored by Fermi Forward Discovery Group, LLC under Contract No. 89243024CSC000002 with the U.S. Department of Energy, Office of Science, Office of High Energy Physics. This work was performed in part at the Aspen Center for Physics, with support for BA by a grant from the Simons Foundation (1161654,Troyer).

\appendix


\bibliographystyle{utphys}
\bibliography{ref}

\providecommand{\href}[2]{#2}\begingroup\raggedright\begin{thebibliography}{10}

\bibitem{Buchmuller:1985jz}
W.~Buchmuller and D.~Wyler, ``{Effective Lagrangian Analysis of New Interactions and Flavor Conservation},''
\href{http://dx.doi.org/10.1016/0550-3213(86)90262-2}{{\em Nucl.Phys.} {\bf B268} (1986)  621--653}.

\bibitem{Grzadkowski:2010es}
B.~Grzadkowski, M.~Iskrzynski, M.~Misiak, and J.~Rosiek, ``{Dimension-Six Terms in the Standard Model Lagrangian},'' \href{http://dx.doi.org/10.1007/JHEP10(2010)085}{{\em JHEP} {\bf 10} (2010)  085}, \href{http://arxiv.org/abs/1008.4884}{{\tt arXiv:1008.4884 [hep-ph]}}.

\bibitem{Brivio:2017vri}
I.~Brivio and M.~Trott, ``{The Standard Model as an Effective Field Theory},'' \href{http://dx.doi.org/10.1016/j.physrep.2018.11.002}{{\em Phys. Rept.} {\bf 793} (2019)  1--98},
\href{http://arxiv.org/abs/1706.08945}{{\tt arXiv:1706.08945 [hep-ph]}}.

\bibitem{Araz:2020zyh}
J.~Y. Araz, S.~Banerjee, R.~S. Gupta, and M.~Spannowsky, ``{Precision SMEFT bounds from the VBF Higgs at high transverse momentum},'' \href{http://dx.doi.org/10.1007/JHEP04(2021)125}{{\em JHEP} {\bf 04} (2021)  125}, \href{http://arxiv.org/abs/2011.03555}{{\tt arXiv:2011.03555 [hep-ph]}}.

\bibitem{Contino:2013kra}
R.~Contino, M.~Ghezzi, C.~Grojean, M.~Muhlleitner, and M.~Spira, ``{Effective Lagrangian for a light Higgs-like scalar},'' \href{http://dx.doi.org/10.1007/JHEP07(2013)035}{{\em JHEP} {\bf 07} (2013)  035},
\href{http://arxiv.org/abs/1303.3876}{{\tt arXiv:1303.3876 [hep-ph]}}.

\bibitem{Falkowski:2014tna}
A.~Falkowski and F.~Riva, ``{Model-independent precision constraints on dimension-6 operators},'' \href{http://dx.doi.org/10.1007/JHEP02(2015)039}{{\em JHEP} {\bf 02} (2015)  039},
\href{http://arxiv.org/abs/1411.0669}{{\tt arXiv:1411.0669 [hep-ph]}}.

\bibitem{Englert:2014cva}
C.~Englert and M.~Spannowsky, ``{Effective Theories and Measurements at Colliders},'' \href{http://dx.doi.org/10.1016/j.physletb.2014.11.035}{{\em Phys. Lett.} {\bf B740} (2015)  8--15},
\href{http://arxiv.org/abs/1408.5147}{{\tt arXiv:1408.5147 [hep-ph]}}.

\bibitem{Gupta:2014rxa}
R.~S. Gupta, A.~Pomarol, and F.~Riva, ``{BSM Primary Effects},'' \href{http://dx.doi.org/10.1103/PhysRevD.91.035001}{{\em Phys. Rev.} {\bf D91} (2015) no.~3, 035001},
\href{http://arxiv.org/abs/1405.0181}{{\tt arXiv:1405.0181 [hep-ph]}}.

\bibitem{Amar:2014fpa}
G.~Amar, S.~Banerjee, S.~von Buddenbrock, A.~S. Cornell, T.~Mandal, B.~Mellado, and B.~Mukhopadhyaya, ``{Exploration of the tensor structure of the Higgs boson coupling to weak bosons in e$^{+}$ e$^{−}$ collisions},'' \href{http://dx.doi.org/10.1007/JHEP02(2015)128}{{\em JHEP} {\bf 02} (2015)  128},
\href{http://arxiv.org/abs/1405.3957}{{\tt arXiv:1405.3957 [hep-ph]}}.

\bibitem{Buschmann:2014sia}
M.~Buschmann, D.~Goncalves, S.~Kuttimalai, M.~Schonherr, F.~Krauss, and T.~Plehn, ``{Mass Effects in the Higgs-Gluon Coupling: Boosted vs Off-Shell Production},'' \href{http://dx.doi.org/10.1007/JHEP02(2015)038}{{\em JHEP} {\bf 02} (2015)  038},
\href{http://arxiv.org/abs/1410.5806}{{\tt arXiv:1410.5806 [hep-ph]}}.

\bibitem{Craig:2014una}
N.~Craig, M.~Farina, M.~McCullough, and M.~Perelstein, ``{Precision Higgsstrahlung as a Probe of New Physics},'' \href{http://dx.doi.org/10.1007/JHEP03(2015)146}{{\em JHEP} {\bf 03} (2015)  146},
\href{http://arxiv.org/abs/1411.0676}{{\tt arXiv:1411.0676 [hep-ph]}}.

\bibitem{Ellis:2014dva}
J.~Ellis, V.~Sanz, and T.~You, ``{Complete Higgs Sector Constraints on Dimension-6 Operators},'' \href{http://dx.doi.org/10.1007/JHEP07(2014)036}{{\em JHEP} {\bf 07} (2014)  036},
\href{http://arxiv.org/abs/1404.3667}{{\tt arXiv:1404.3667 [hep-ph]}}.

\bibitem{Ellis:2014jta}
J.~Ellis, V.~Sanz, and T.~You, ``{The Effective Standard Model after LHC Run I},'' \href{http://dx.doi.org/10.1007/JHEP03(2015)157}{{\em JHEP} {\bf 03} (2015)  157},
\href{http://arxiv.org/abs/1410.7703}{{\tt arXiv:1410.7703 [hep-ph]}}.

\bibitem{Banerjee:2015bla}
S.~Banerjee, T.~Mandal, B.~Mellado, and B.~Mukhopadhyaya, ``{Cornering dimension-6 $HVV$ interactions at high luminosity LHC: the role of event ratios},'' \href{http://dx.doi.org/10.1007/JHEP09(2015)057}{{\em JHEP} {\bf 09} (2015)  057},
\href{http://arxiv.org/abs/1505.00226}{{\tt arXiv:1505.00226 [hep-ph]}}.

\bibitem{Englert:2015hrx}
C.~Englert, R.~Kogler, H.~Schulz, and M.~Spannowsky, ``{Higgs coupling measurements at the LHC},'' \href{http://dx.doi.org/10.1140/epjc/s10052-016-4227-1}{{\em Eur. Phys. J.} {\bf C76} (2016) no.~7, 393},
\href{http://arxiv.org/abs/1511.05170}{{\tt arXiv:1511.05170 [hep-ph]}}.

\bibitem{Contino:2016jqw}
R.~Contino, A.~Falkowski, F.~Goertz, C.~Grojean, and F.~Riva, ``{On the Validity of the Effective Field Theory Approach to SM Precision Tests},'' \href{http://dx.doi.org/10.1007/JHEP07(2016)144}{{\em JHEP} {\bf 07} (2016)  144},
\href{http://arxiv.org/abs/1604.06444}{{\tt arXiv:1604.06444 [hep-ph]}}.

\bibitem{Biekotter:2016ecg}
A.~Biek{\"o}tter, J.~Brehmer, and T.~Plehn, ``{Extending the limits of Higgs effective theory},'' \href{http://dx.doi.org/10.1103/PhysRevD.94.055032}{{\em Phys. Rev.} {\bf D94} (2016) no.~5, 055032},
\href{http://arxiv.org/abs/1602.05202}{{\tt arXiv:1602.05202 [hep-ph]}}.

\bibitem{Barklow:2017suo}
T.~Barklow, K.~Fujii, S.~Jung, R.~Karl, J.~List, T.~Ogawa, M.~E. Peskin, and J.~Tian, ``{Improved Formalism for Precision Higgs Coupling Fits},'' \href{http://dx.doi.org/10.1103/PhysRevD.97.053003}{{\em Phys. Rev.} {\bf D97} (2018) no.~5, 053003},
\href{http://arxiv.org/abs/1708.08912}{{\tt arXiv:1708.08912 [hep-ph]}}.

\bibitem{Barklow:2017awn}
T.~Barklow, K.~Fujii, S.~Jung, M.~E. Peskin, and J.~Tian, ``{Model-Independent Determination of the Triple Higgs Coupling at e+e- Colliders},'' \href{http://dx.doi.org/10.1103/PhysRevD.97.053004}{{\em Phys. Rev.} {\bf D97} (2018) no.~5, 053004},
\href{http://arxiv.org/abs/1708.09079}{{\tt arXiv:1708.09079 [hep-ph]}}.

\bibitem{Englert:2017aqb}
C.~Englert, R.~Kogler, H.~Schulz, and M.~Spannowsky, ``{Higgs characterisation in the presence of theoretical uncertainties and invisible decays},'' \href{http://dx.doi.org/10.1140/epjc/s10052-017-5366-8}{{\em Eur. Phys. J.} {\bf C77} (2017) no.~11, 789},
\href{http://arxiv.org/abs/1708.06355}{{\tt arXiv:1708.06355 [hep-ph]}}.

\bibitem{banerjee1}
S.~Banerjee, C.~Englert, R.~S. Gupta, and M.~Spannowsky, ``{Probing Electroweak Precision Physics via boosted Higgs-strahlung at the LHC},'' \href{http://dx.doi.org/10.1103/PhysRevD.98.095012}{{\em Phys. Rev. D} {\bf 98} (2018) no.~9, 095012}, \href{http://arxiv.org/abs/1807.01796}{{\tt arXiv:1807.01796 [hep-ph]}}.

\bibitem{Grojean:2018dqj}
C.~Grojean, M.~Montull, and M.~Riembau, ``Diboson at the lhc vs lep,'' \href{http://dx.doi.org/10.1007/JHEP03(2019)020}{{\em JHEP} {\bf 03} (2019)  020}, \href{http://arxiv.org/abs/1810.05149}{{\tt arXiv:1810.05149 [hep-ph]}}.

\bibitem{Biekotter:2018rhp}
A.~Biekoetter, T.~Corbett, and T.~Plehn, ``{The Gauge-Higgs Legacy of the LHC Run II},'' \href{http://dx.doi.org/10.21468/SciPostPhys.6.6.064}{{\em SciPost Phys.} {\bf 6} (2019) no.~6, 064}, \href{http://arxiv.org/abs/1812.07587}{{\tt arXiv:1812.07587 [hep-ph]}}.

\bibitem{Goncalves:2018ptp}
D.~Goncalves and J.~Nakamura, ``{Boosting the $H\to$ invisibles searches with $Z$ boson polarization},'' \href{http://dx.doi.org/10.1103/PhysRevD.99.055021}{{\em Phys. Rev.} {\bf D99} (2019) no.~5, 055021},
\href{http://arxiv.org/abs/1809.07327}{{\tt arXiv:1809.07327 [hep-ph]}}.

\bibitem{Gomez-Ambrosio:2018pnl}
R.~Gomez-Ambrosio, ``{Studies of Dimension-Six EFT effects in Vector Boson Scattering},'' \href{http://dx.doi.org/10.1140/epjc/s10052-019-6893-2}{{\em Eur. Phys. J. C} {\bf 79} (2019) no.~5, 389}, \href{http://arxiv.org/abs/1809.04189}{{\tt arXiv:1809.04189 [hep-ph]}}.

\bibitem{Freitas:2019hbk}
F.~F. Freitas, C.~K. Khosa, and V.~Sanz, ``{Exploring SMEFT in VH with Machine Learning},''
\href{http://arxiv.org/abs/1902.05803}{{\tt arXiv:1902.05803 [hep-ph]}}.

\bibitem{Banerjee:2019pks}
S.~Banerjee, R.~S. Gupta, J.~Y. Reiness, and M.~Spannowsky, ``{Resolving the tensor structure of the Higgs coupling to $Z$-bosons via Higgs-strahlung},'' \href{http://dx.doi.org/10.1103/PhysRevD.100.115004}{{\em Phys. Rev.} {\bf D100} (2019) no.~11, 115004},
\href{http://arxiv.org/abs/1905.02728}{{\tt arXiv:1905.02728 [hep-ph]}}.

\bibitem{Banerjee:2019twi}
S.~Banerjee, R.~S. Gupta, J.~Y. Reiness, S.~Seth, and M.~Spannowsky, ``{Towards the ultimate differential SMEFT analysis},'' \href{http://dx.doi.org/10.1007/JHEP09(2020)170}{{\em JHEP} {\bf 09} (2020)  170}, \href{http://arxiv.org/abs/1912.07628}{{\tt arXiv:1912.07628 [hep-ph]}}.

\bibitem{Biekotter:2020flu}
A.~Biek\"otter, R.~Gomez-Ambrosio, P.~Gregg, F.~Krauss, and M.~Sch\"onherr, ``{Constraining SMEFT operators with associated $h\gamma$ production in Weak Boson Fusion},'' \href{http://arxiv.org/abs/2003.06379}{{\tt arXiv:2003.06379 [hep-ph]}}.

\bibitem{Ellis:2018gqa}
J.~Ellis, C.~W. Murphy, V.~Sanz, and T.~You, ``{Updated Global SMEFT Fit to Higgs, Diboson and Electroweak Data},'' \href{http://dx.doi.org/10.1007/JHEP06(2018)146}{{\em JHEP} {\bf 06} (2018)  146}, \href{http://arxiv.org/abs/1803.03252}{{\tt arXiv:1803.03252 [hep-ph]}}.

\bibitem{Ellis:2020unq}
J.~Ellis, M.~Madigan, K.~Mimasu, V.~Sanz, and T.~You, ``{Top, Higgs, Diboson and Electroweak Fit to the Standard Model Effective Field Theory},'' \href{http://dx.doi.org/10.1007/JHEP04(2021)279}{{\em JHEP} {\bf 04} (2021)  279}, \href{http://arxiv.org/abs/2012.02779}{{\tt arXiv:2012.02779 [hep-ph]}}.

\bibitem{Bartocci:2023nvp}
R.~Bartocci, A.~Biek\"otter, and T.~Hurth, ``{A global analysis of the SMEFT under the minimal MFV assumption},'' \href{http://dx.doi.org/10.1007/JHEP05(2024)074}{{\em JHEP} {\bf 05} (2024)  074}, \href{http://arxiv.org/abs/2311.04963}{{\tt arXiv:2311.04963 [hep-ph]}}.

\bibitem{Giani:2023gfq}
T.~Giani, G.~Magni, and J.~Rojo, ``{SMEFiT: a flexible toolbox for global interpretations of particle physics data with effective field theories},'' \href{http://dx.doi.org/10.1140/epjc/s10052-023-11534-7}{{\em Eur. Phys. J. C} {\bf 83} (2023) no.~5, 393}, \href{http://arxiv.org/abs/2302.06660}{{\tt arXiv:2302.06660 [hep-ph]}}.

\bibitem{Dedes:2017zog}
A.~Dedes, W.~Materkowska, M.~Paraskevas, J.~Rosiek, and K.~Suxho, ``{Feynman rules for the Standard Model Effective Field Theory in $R_{\xi}$ -gauges},'' \href{http://dx.doi.org/10.1007/JHEP06(2017)143}{{\em JHEP} {\bf 06} (2017)  143}, \href{http://arxiv.org/abs/1704.03888}{{\tt arXiv:1704.03888 [hep-ph]}}.

\bibitem{Dedes:2023zws}
A.~Dedes, J.~Rosiek, M.~Ryczkowski, K.~Suxho, and L.~Trifyllis, ``{SmeftFR v3 \textendash{} Feynman rules generator for the Standard Model Effective Field Theory},'' \href{http://dx.doi.org/10.1016/j.cpc.2023.108943}{{\em Comput. Phys. Commun.} {\bf 294} (2024)  108943}, \href{http://arxiv.org/abs/2302.01353}{{\tt arXiv:2302.01353 [hep-ph]}}.

\bibitem{Henning:2015alf}
B.~Henning, X.~Lu, T.~Melia, and H.~Murayama, ``{2, 84, 30, 993, 560, 15456, 11962, 261485, ...: Higher dimension operators in the SM EFT},'' \href{http://dx.doi.org/10.1007/JHEP08(2017)016}{{\em JHEP} {\bf 08} (2017)  016}, \href{http://arxiv.org/abs/1512.03433}{{\tt arXiv:1512.03433 [hep-ph]}}. [Erratum: JHEP 09, 019 (2019)].

\bibitem{Assi:2023zid}
B.~Assi, A.~Helset, A.~V. Manohar, J.~Pag\`es, and C.-H. Shen, ``{Fermion geometry and the renormalization of the Standard Model Effective Field Theory},'' \href{http://dx.doi.org/10.1007/JHEP11(2023)201}{{\em JHEP} {\bf 11} (2023)  201}, \href{http://arxiv.org/abs/2307.03187}{{\tt arXiv:2307.03187 [hep-ph]}}.

\bibitem{Chala:2021pll}
M.~Chala, G.~Guedes, M.~Ramos, and J.~Santiago, ``{Towards the renormalisation of the Standard Model effective field theory to dimension eight: Bosonic interactions I},'' \href{http://dx.doi.org/10.21468/SciPostPhys.11.3.065}{{\em SciPost Phys.} {\bf 11} (2021)  065}, \href{http://arxiv.org/abs/2106.05291}{{\tt arXiv:2106.05291 [hep-ph]}}.

\bibitem{DasBakshi:2022mwk}
S.~Das~Bakshi, M.~Chala, A.~D\'\i{}az-Carmona, and G.~Guedes, ``{Towards the renormalisation of the Standard Model effective field theory to dimension eight: bosonic interactions II},'' \href{http://dx.doi.org/10.1140/epjp/s13360-022-03194-5}{{\em Eur. Phys. J. Plus} {\bf 137} (2022) no.~8, 973}, \href{http://arxiv.org/abs/2205.03301}{{\tt arXiv:2205.03301 [hep-ph]}}.

\bibitem{Bakshi:2024wzz}
S.~D. Bakshi, M.~Chala, A.~D\'\i{}az-Carmona, Z.~Ren, and F.~Vilches, ``{Renormalization of the SMEFT to dimension eight: Fermionic interactions I},'' \href{http://dx.doi.org/10.1007/JHEP12(2024)214}{{\em JHEP} {\bf 12} (2025)  214}, \href{http://arxiv.org/abs/2409.15408}{{\tt arXiv:2409.15408 [hep-ph]}}.

\bibitem{Boughezal:2024zqa}
R.~Boughezal, Y.~Huang, and F.~Petriello, ``{Renormalization-group running of dimension-8 four-fermion operators in the SMEFT},'' \href{http://dx.doi.org/10.1103/PhysRevD.110.116015}{{\em Phys. Rev. D} {\bf 110} (2024) no.~11, 116015}, \href{http://arxiv.org/abs/2408.15378}{{\tt arXiv:2408.15378 [hep-ph]}}.

\bibitem{Helset:2022pde}
A.~Helset, E.~E. Jenkins, and A.~V. Manohar, ``{Renormalization of the Standard Model Effective Field Theory from geometry},'' \href{http://dx.doi.org/10.1007/JHEP02(2023)063}{{\em JHEP} {\bf 02} (2023)  063}, \href{http://arxiv.org/abs/2212.03253}{{\tt arXiv:2212.03253 [hep-ph]}}.

\bibitem{Assi:2024zap}
B.~Assi and A.~Martin, ``{Energy-enhanced dimension eight SMEFT effects in VBF Higgs production},'' \href{http://dx.doi.org/10.1007/JHEP02(2025)029}{{\em JHEP} {\bf 02} (2025)  029}, \href{http://arxiv.org/abs/2410.21563}{{\tt arXiv:2410.21563 [hep-ph]}}.

\bibitem{Kim:2022amu}
T.~Kim and A.~Martin, ``{Monolepton production in SMEFT to $ \mathcal{O} $(1/\ensuremath{\Lambda}$^{4}$) and beyond},'' \href{http://dx.doi.org/10.1007/JHEP09(2022)124}{{\em JHEP} {\bf 09} (2022)  124}, \href{http://arxiv.org/abs/2203.11976}{{\tt arXiv:2203.11976 [hep-ph]}}.

\bibitem{Martin:2023tvi}
A.~Martin, ``{A case study of SMEFT $ \mathcal{O}\left(1/{\Lambda}^4\right) $ effects in diboson processes: pp \textrightarrow{} W$^{±}$(\ensuremath{\ell}$^{±}$\ensuremath{\nu})\ensuremath{\gamma}},'' \href{http://dx.doi.org/10.1007/JHEP05(2024)223}{{\em JHEP} {\bf 05} (2024)  223}, \href{http://arxiv.org/abs/2312.09867}{{\tt arXiv:2312.09867 [hep-ph]}}.

\bibitem{Corbett:2023yhk}
T.~Corbett and A.~Martin, ``{Higgs associated production with a vector decaying to two fermions in the geoSMEFT},'' \href{http://arxiv.org/abs/2306.00053}{{\tt arXiv:2306.00053 [hep-ph]}}.

\bibitem{Degrande:2023iob}
C.~Degrande and H.-L. Li, ``{Impact of dimension-8 SMEFT operators on diboson productions},'' \href{http://dx.doi.org/10.1007/JHEP06(2023)149}{{\em JHEP} {\bf 06} (2023)  149}, \href{http://arxiv.org/abs/2303.10493}{{\tt arXiv:2303.10493 [hep-ph]}}.

\bibitem{Helset:2020yio}
A.~Helset, A.~Martin, and M.~Trott, ``{The Geometric Standard Model Effective Field Theory},'' \href{http://dx.doi.org/10.1007/JHEP03(2020)163}{{\em JHEP} {\bf 03} (2020)  163}, \href{http://arxiv.org/abs/2001.01453}{{\tt arXiv:2001.01453 [hep-ph]}}.

\bibitem{deBlas:2014mba}
J.~de~Blas, M.~Chala, M.~Perez-Victoria, and J.~Santiago, ``{Observable Effects of General New Scalar Particles},'' \href{http://dx.doi.org/10.1007/JHEP04(2015)078}{{\em JHEP} {\bf 04} (2015)  078}, \href{http://arxiv.org/abs/1412.8480}{{\tt arXiv:1412.8480 [hep-ph]}}.

\bibitem{deBlas:2017xtg}
J.~de~Blas, J.~C. Criado, M.~Perez-Victoria, and J.~Santiago, ``{Effective description of general extensions of the Standard Model: the complete tree-level dictionary},'' \href{http://dx.doi.org/10.1007/JHEP03(2018)109}{{\em JHEP} {\bf 03} (2018)  109}, \href{http://arxiv.org/abs/1711.10391}{{\tt arXiv:1711.10391 [hep-ph]}}.

\bibitem{Carmona:2021xtq}
A.~Carmona, A.~Lazopoulos, P.~Olgoso, and J.~Santiago, ``{Matchmakereft: automated tree-level and one-loop matching},'' \href{http://dx.doi.org/10.21468/SciPostPhys.12.6.198}{{\em SciPost Phys.} {\bf 12} (2022) no.~6, 198}, \href{http://arxiv.org/abs/2112.10787}{{\tt arXiv:2112.10787 [hep-ph]}}.

\bibitem{Fuentes-Martin:2022jrf}
J.~Fuentes-Mart\'\i{}n, M.~K\"onig, J.~Pag\`es, A.~E. Thomsen, and F.~Wilsch, ``{A proof of concept for matchete: an automated tool for matching effective theories},'' \href{http://dx.doi.org/10.1140/epjc/s10052-023-11726-1}{{\em Eur. Phys. J. C} {\bf 83} (2023) no.~7, 662}, \href{http://arxiv.org/abs/2212.04510}{{\tt arXiv:2212.04510 [hep-ph]}}.

\bibitem{Li:2023cwy}
X.-X. Li, Z.~Ren, and J.-H. Yub, ``{Complete tree-level dictionary between simplified BSM models and SMEFT d\ensuremath{\leq}7 operators},'' \href{http://dx.doi.org/10.1103/PhysRevD.109.095041}{{\em Phys. Rev. D} {\bf 109} (2024) no.~9, 095041}, \href{http://arxiv.org/abs/2307.10380}{{\tt arXiv:2307.10380 [hep-ph]}}.

\bibitem{Hays:2020scx}
C.~Hays, A.~Helset, A.~Martin, and M.~Trott, ``{Exact SMEFT formulation and expansion to $\mathcal{O}(v^4/\Lambda^4)$},'' \href{http://dx.doi.org/10.1007/JHEP11(2020)087}{{\em JHEP} {\bf 11} (2020)  087}, \href{http://arxiv.org/abs/2007.00565}{{\tt arXiv:2007.00565 [hep-ph]}}.

\bibitem{Corbett:2021eux}
T.~Corbett, A.~Helset, A.~Martin, and M.~Trott, ``{EWPD in the SMEFT to dimension eight},'' \href{http://dx.doi.org/10.1007/JHEP06(2021)076}{{\em JHEP} {\bf 06} (2021)  076}, \href{http://arxiv.org/abs/2102.02819}{{\tt arXiv:2102.02819 [hep-ph]}}.

\bibitem{Shadmi:2018xan}
Y.~Shadmi and Y.~Weiss, ``{Effective Field Theory Amplitudes the On-Shell Way: Scalar and Vector Couplings to Gluons},'' \href{http://dx.doi.org/10.1007/JHEP02(2019)165}{{\em JHEP} {\bf 02} (2019)  165}, \href{http://arxiv.org/abs/1809.09644}{{\tt arXiv:1809.09644 [hep-ph]}}.

\bibitem{Ma:2019gtx}
T.~Ma, J.~Shu, and M.-L. Xiao, ``{Standard model effective field theory from on-shell amplitudes*},'' \href{http://dx.doi.org/10.1088/1674-1137/aca200}{{\em Chin. Phys. C} {\bf 47} (2023) no.~2, 023105}, \href{http://arxiv.org/abs/1902.06752}{{\tt arXiv:1902.06752 [hep-ph]}}.

\bibitem{Aoude:2019tzn}
R.~Aoude and C.~S. Machado, ``{The Rise of SMEFT On-shell Amplitudes},'' \href{http://dx.doi.org/10.1007/JHEP12(2019)058}{{\em JHEP} {\bf 12} (2019)  058}, \href{http://arxiv.org/abs/1905.11433}{{\tt arXiv:1905.11433 [hep-ph]}}.

\bibitem{Durieux:2020gip}
G.~Durieux, T.~Kitahara, C.~S. Machado, Y.~Shadmi, and Y.~Weiss, ``{Constructing massive on-shell contact terms},'' \href{http://dx.doi.org/10.1007/JHEP12(2020)175}{{\em JHEP} {\bf 12} (2020)  175}, \href{http://arxiv.org/abs/2008.09652}{{\tt arXiv:2008.09652 [hep-ph]}}.

\bibitem{AccettulliHuber:2021uoa}
M.~Accettulli~Huber and S.~De~Angelis, ``{Standard Model EFTs via on-shell methods},'' \href{http://dx.doi.org/10.1007/JHEP11(2021)221}{{\em JHEP} {\bf 11} (2021)  221}, \href{http://arxiv.org/abs/2108.03669}{{\tt arXiv:2108.03669 [hep-th]}}.

\bibitem{Arzt_1995}
C.~Arzt, M.~Einhorn, and J.~Wudka, ``Patterns of deviation from the standard model,'' \href{http://dx.doi.org/10.1016/0550-3213(94)00336-d}{{\em Nuclear Physics B} {\bf 433} (1995) no.~1, 41–66}. \url{http://dx.doi.org/10.1016/0550-3213(94)00336-D}.

\bibitem{Buchalla:2022vjp}
G.~Buchalla, G.~Heinrich, C.~M\"uller-Salditt, and F.~Pandler, ``{Loop counting matters in SMEFT},'' \href{http://dx.doi.org/10.21468/SciPostPhys.15.3.088}{{\em SciPost Phys.} {\bf 15} (2023) no.~3, 088}, \href{http://arxiv.org/abs/2204.11808}{{\tt arXiv:2204.11808 [hep-ph]}}.

\bibitem{Craig:2019wmo}
N.~Craig, M.~Jiang, Y.-Y. Li, and D.~Sutherland, ``{Loops and Trees in Generic EFTs},'' \href{http://dx.doi.org/10.1007/JHEP08(2020)086}{{\em JHEP} {\bf 08} (2020)  086}, \href{http://arxiv.org/abs/2001.00017}{{\tt arXiv:2001.00017 [hep-ph]}}.

\bibitem{DAmbrosio:2002vsn}
G.~D'Ambrosio, G.~F. Giudice, G.~Isidori, and A.~Strumia, ``{Minimal flavor violation: An Effective field theory approach},'' \href{http://dx.doi.org/10.1016/S0550-3213(02)00836-2}{{\em Nucl. Phys. B} {\bf 645} (2002)  155--187}, \href{http://arxiv.org/abs/hep-ph/0207036}{{\tt arXiv:hep-ph/0207036}}.

\bibitem{Bishara:2022vsc}
F.~Bishara, P.~Englert, C.~Grojean, G.~Panico, and A.~N. Rossia, ``{Revisiting Vh(\textrightarrow{}$ b\overline{b} $) at the LHC and FCC-hh},'' \href{http://dx.doi.org/10.1007/JHEP06(2023)077}{{\em JHEP} {\bf 06} (2023)  077}, \href{http://arxiv.org/abs/2208.11134}{{\tt arXiv:2208.11134 [hep-ph]}}.

\bibitem{Celada:2024mcf}
E.~Celada, T.~Giani, J.~ter Hoeve, L.~Mantani, J.~Rojo, A.~N. Rossia, M.~O.~A. Thomas, and E.~Vryonidou, ``{Mapping the SMEFT at high-energy colliders: from LEP and the (HL-)LHC to the FCC-ee},'' \href{http://dx.doi.org/10.1007/JHEP09(2024)091}{{\em JHEP} {\bf 09} (2024)  091}, \href{http://arxiv.org/abs/2404.12809}{{\tt arXiv:2404.12809 [hep-ph]}}.

\bibitem{Hays:2018zze}
C.~Hays, A.~Martin, V.~Sanz, and J.~Setford, ``{On the impact of dimension-eight SMEFT operators on Higgs measurements},'' \href{http://dx.doi.org/10.1007/JHEP02(2019)123}{{\em JHEP} {\bf 02} (2019)  123}, \href{http://arxiv.org/abs/1808.00442}{{\tt arXiv:1808.00442 [hep-ph]}}.

\bibitem{Elvang:2013cua}
H.~Elvang and Y.-t. Huang, ``{Scattering Amplitudes},'' \href{http://arxiv.org/abs/1308.1697}{{\tt arXiv:1308.1697 [hep-th]}}.

\bibitem{Arganda:2018ftn}
E.~Arganda, C.~Garcia-Garcia, and M.~J. Herrero, ``{Probing the Higgs self-coupling through double Higgs production in vector boson scattering at the LHC},'' \href{http://dx.doi.org/10.1016/j.nuclphysb.2019.114687}{{\em Nucl. Phys. B} {\bf 945} (2019)  114687}, \href{http://arxiv.org/abs/1807.09736}{{\tt arXiv:1807.09736 [hep-ph]}}.

\bibitem{Cepeda:2019klc}
M.~Cepeda {\em et al.}, ``{Report from Working Group 2}: {Higgs Physics at the HL-LHC and HE-LHC},'' \href{http://dx.doi.org/10.23731/CYRM-2019-007.221}{{\em CERN Yellow Rep. Monogr.} {\bf 7} (2019)  221--584}, \href{http://arxiv.org/abs/1902.00134}{{\tt arXiv:1902.00134 [hep-ph]}}.

\bibitem{Chen:2021rid}
J.~Chen, C.-T. Lu, and Y.~Wu, ``{Measuring Higgs boson self-couplings with 2 \textrightarrow{} 3 VBS processes},'' \href{http://dx.doi.org/10.1007/JHEP10(2021)099}{{\em JHEP} {\bf 10} (2021)  099}, \href{http://arxiv.org/abs/2105.11500}{{\tt arXiv:2105.11500 [hep-ph]}}.

\bibitem{Gomez-Ambrosio:2022qsi}
R.~G\'omez-Ambrosio, F.~J. Llanes-Estrada, A.~Salas-Bern\'ardez, and J.~J. Sanz-Cillero, ``{Distinguishing electroweak EFTs with WLWL\textrightarrow{}n\texttimes{}h},'' \href{http://dx.doi.org/10.1103/PhysRevD.106.053004}{{\em Phys. Rev. D} {\bf 106} (2022) no.~5, 053004}, \href{http://arxiv.org/abs/2204.01763}{{\tt arXiv:2204.01763 [hep-ph]}}.

\bibitem{CMS:2022gjd}
{\bf CMS} Collaboration, A.~Tumasyan {\em et al.}, ``{Search for Nonresonant Pair Production of Highly Energetic Higgs Bosons Decaying to Bottom Quarks},'' \href{http://dx.doi.org/10.1103/PhysRevLett.131.041803}{{\em Phys. Rev. Lett.} {\bf 131} (2023) no.~4, 041803}, \href{http://arxiv.org/abs/2205.06667}{{\tt arXiv:2205.06667 [hep-ex]}}.

\bibitem{Cappati:2022skp}
A.~Cappati, R.~Covarelli, P.~Torrielli, and M.~Zaro, ``{Sensitivity to new physics in final states with multiple gauge and Higgs bosons},'' \href{http://dx.doi.org/10.1007/JHEP09(2022)038}{{\em JHEP} {\bf 09} (2022)  038}, \href{http://arxiv.org/abs/2205.15959}{{\tt arXiv:2205.15959 [hep-ph]}}.

\bibitem{Delgado:2023ynh}
R.~L. Delgado, R.~G\'omez-Ambrosio, J.~Mart\'\i{}nez-Mart\'\i{}n, A.~Salas-Bern\'ardez, and J.~J. Sanz-Cillero, ``{Production of two, three, and four Higgs bosons: where SMEFT and HEFT depart},'' \href{http://dx.doi.org/10.1007/JHEP03(2024)037}{{\em JHEP} {\bf 03} (2024)  037}, \href{http://arxiv.org/abs/2311.04280}{{\tt arXiv:2311.04280 [hep-ph]}}.

\end{thebibliography}\endgroup

\end{document}